\pdfoutput=1
\documentclass[acmsmall,screen]{acmart}

\setcopyright{rightsretained}
\acmPrice{}
\acmDOI{10.1145/3290386}
\acmYear{2019}
\copyrightyear{2019}
\acmJournal{PACMPL}
\acmVolume{3}
\acmNumber{POPL}
\acmArticle{73}
\acmMonth{1}

\usepackage{mathtools}
\usepackage[skip=4pt]{caption}
\usepackage{subcaption}
\usepackage{bold-extra}
\usepackage{booktabs}
\usepackage{enumitem}
\usepackage{wrapfig}

\usepackage{mdframed}
\usepackage{listings}

\mdfsetup{skipbelow=4pt,skipabove=6pt,leftmargin=20pt,rightmargin=10pt,align=left,usetwoside=false}

\mdfdefinestyle{listingstyle}{
  backgroundcolor=black!2,
  linewidth=2pt,linecolor=black!20,
  outerlinewidth=5pt,outerlinecolor=black,
  rightline=false,topline=false,bottomline=false,
  innerleftmargin=8pt,innerrightmargin=4pt,innertopmargin=0pt,innerbottommargin=0pt,
}

\surroundwithmdframed[style=listingstyle]{lstlisting}

\makeatletter 
\lst@AddToHook{OnEmptyLine}{\vspace{\dimexpr-\baselineskip+2\smallskipamount}} 
\makeatother 

\usepackage{algorithm}
\usepackage[noend]{algpseudocode}
\usepackage{algorithmicx}
\algblockdefx{RepeatUntilTimeout}{EndRepeat}{\textbf{repeat until timeout}}{}
\algblockdefx{RepeatUntil}{EndRepeat}{\textbf{repeat until }}{}
\algtext*{EndRepeat}
\usepackage{float}
\newfloat{algorithm}{t}{}

\algnewcommand\algorithmicinput{\textbf{Input:}}
\algnewcommand\Input{\item[\algorithmicinput]}
\algnewcommand\algorithmicoutput{\textbf{Output:}}
\algnewcommand\Output{\item[\algorithmicoutput]}
\makeatletter
\algnewcommand{\LineComment}[1]{\Statex \hskip\ALG@thistlm \(\triangleright\) #1}
\makeatother

\usepackage{etoolbox}
\usepackage{tikz}
\usetikzlibrary{tikzmark}
\usetikzlibrary{calc}

\newcommand{\ALGtikzmarkcolor}{lightgray}
\newcommand{\ALGtikzmarkextraindent}{4pt}
\newcommand{\ALGtikzmarkverticaloffsetstart}{-.7ex}
\newcommand{\ALGtikzmarkverticaloffsetend}{-.5ex}
\makeatletter
\newcounter{ALG@tikzmark@tempcnta}

\newcommand\ALG@tikzmark@start{%
    \global\let\ALG@tikzmark@last\ALG@tikzmark@starttext%
    \expandafter\edef\csname ALG@tikzmark@\theALG@nested\endcsname{\theALG@tikzmark@tempcnta}%
    \tikzmark{ALG@tikzmark@start@\csname ALG@tikzmark@\theALG@nested\endcsname}%
    \addtocounter{ALG@tikzmark@tempcnta}{1}%
}

\def\ALG@tikzmark@starttext{start}
\newcommand\ALG@tikzmark@end{%
    \ifx\ALG@tikzmark@last\ALG@tikzmark@starttext
    \else
        \tikzmark{ALG@tikzmark@end@\csname ALG@tikzmark@\theALG@nested\endcsname}%
        \tikz[overlay,remember picture] \draw[\ALGtikzmarkcolor] let \p{S}=($(pic cs:ALG@tikzmark@start@\csname ALG@tikzmark@\theALG@nested\endcsname)+(\ALGtikzmarkextraindent,\ALGtikzmarkverticaloffsetstart)$), \p{E}=($(pic cs:ALG@tikzmark@end@\csname ALG@tikzmark@\theALG@nested\endcsname)+(\ALGtikzmarkextraindent,\ALGtikzmarkverticaloffsetend)$) in (\x{S},\y{S})--(\x{S},\y{E});%
    \fi
    \gdef\ALG@tikzmark@last{end}%
}

\apptocmd{\ALG@beginblock}{\ALG@tikzmark@start}{}{\errmessage{failed to patch}}
\pretocmd{\ALG@endblock}{\ALG@tikzmark@end}{}{\errmessage{failed to patch}}
\makeatother

\newcommand{\code}[1]{\texttt{\color{blue!50!black}\smaller{#1}}}
\newcommand{\blackcirc}{{\color{black}\circ}}
\newcommand{\angelic}{\color{blue!50!black}\texttt{\bfseries<ANGELIC>}} 
\newcommand{\frangel}{\textsc{FrAngel}}
\newcommand{\sypet}{\textsc{SyPet}}

\newtheoremstyle{mydef}{6pt}{6pt}{\itshape}{}{\bfseries}{}{1em}{\thmname{#1}~\thmnumber{#2}:\thmnote{~#3.}}
\theoremstyle{mydef}
\newtheorem{defn}{Definition}

\begin{document}

\title[\frangel: Component-Based Synthesis with Control Structures]{\frangel: Component-Based Synthesis \\ with Control Structures}

\author{Kensen Shi}
\email{kensens@cs.stanford.edu}
\affiliation{
    \department{Department of Computer Science}
    \institution{Stanford University}
    \state{California}
    \country{USA}
}
\author{Jacob Steinhardt}
\email{jsteinhardt@cs.stanford.edu}
\affiliation{
    \department{Department of Computer Science}
    \institution{Stanford University}
    \state{California}
    \country{USA}
}
\author{Percy Liang}
\email{pliang@cs.stanford.edu}
\affiliation{
    \department{Department of Computer Science}
    \institution{Stanford University}
    \state{California}
    \country{USA}
}

\begin{abstract}

    In component-based program synthesis, the synthesizer generates a program given a library of components (functions).
    Existing component-based synthesizers have difficulty synthesizing loops and other control structures, and they
    often require formal specifications of the components, which can be expensive to generate. We present \frangel, a
    new approach to component-based synthesis that can synthesize short Java functions with control structures when
    given a desired signature, a set of input-output examples, and a collection of libraries (without formal
    specifications). \frangel\ aims to discover programs with many distinct behaviors by combining two main ideas.
    First, it \emph{mines code fragments} from partially-successful programs that only pass some of the examples. These
    extracted fragments are often useful for synthesis due to a property that we call \emph{special-case similarity}.
    Second, \frangel\ uses \emph{angelic conditions} as placeholders for control structure conditions and optimistically
    evaluates the resulting program sketches. Angelic conditions decompose the synthesis process: \frangel\ first finds
    promising partial programs and later fills in their missing conditions. We demonstrate that \frangel\ can synthesize
    a variety of interesting programs with combinations of control structures within seconds, significantly
    outperforming prior state-of-the-art.

\end{abstract}



\maketitle

\section{Introduction}
\label{sec:intro}

Programmers often browse software libraries to identify useful functions and combine such functions with loops and
conditionals to achieve some desired functionality. Using large libraries is not easy---even experienced programmers can
spend hours identifying the few necessary functions in a library~\cite{jungloid}. \emph{Component-based program
synthesis} aims to ease this aspect of programming by synthesizing programs using a library of \emph{components}
(existing functions).

We focus on synthesis \emph{from examples}, where the user provides input-output examples that the synthesized program
must satisfy. Although the desired program behavior can be specified with logical constraints~\cite{slfp, verification}
or executables~\cite{oracle, mimic}, I/O examples are appealing because they provide an intuitive user experience for
programmers (after all, most unit tests are in fact I/O examples) and even end-users with no programming
knowledge~\cite{flashfill}.

Many works in component-based synthesis use domain-specific knowledge to solve specific classes of problems such as
string manipulation~\cite{flashfill, mlpbe}, geometric constructions~\cite{geometry}, and transformations of
data~\cite{hades, lambda2, morpheus, progfromex}. Similarly, \citet{tds} describe a general framework that accepts an
expert-written domain-specific language (DSL) as input. On the other hand, \citet{slfp, oracle} describe approaches that
are domain-agnostic but require formal specifications of the components. However, few libraries have existing
specifications, and writing a specification for every library component can be prohibitively time-consuming---such
approaches have only been demonstrated in domains with few components and simple specifications (e.g., bit-manipulation
algorithms~\cite{oracle}).

One goal of this work is to perform synthesis from examples without requiring specifications of components or
domain-specific knowledge as in the work described above (this regime was previously considered by \sypet~\cite{sypet}).
To this end, we avoid creating a DSL or requiring one as input. Furthermore, without specifications of components, the
constraint-solving techniques used in \citet{oracle, slfp, morpheus} become infeasible. Like \sypet, our work relies on
blackbox execution rather than reasoning about formal semantics. 

Another goal of our work is to perform synthesis \emph{with control structures}, i.e., loops and conditionals. Loops in
particular have been a common source of difficulty in component-based synthesis \cite{sypet, slfp, oracle, mlpbe}. Some
approaches use specialized loops in a DSL~\cite{flashfill, tds} or looping components (e.g., \code{map} and
\code{filter} operators) \cite{progfromex, lambda2, hades, morpheus} to mitigate this issue in specific domains. Yet,
relying only on such components is insufficient for general-purpose synthesis, evidenced by the fact that programmers
still use loops in everyday code.

We present a system called \frangel\ to tackle domain-agnostic component-based synthesis from examples, using control
structures and libraries without specifications. The key insight is that \emph{progress in a synthesis search is
primarily made by finding programs with distinct meaningful behaviors (i.e., program semantics)}. After all, the goal of
synthesis is not to find one particular program, but rather any program with the desired behavior, possibly among many
acceptable solutions. Furthermore, a meaningful program with nontrivial behavior is composed of simpler ``building
blocks'' that are also meaningful in related ways toward the overall functionality. By identifying and combining
programs with distinct behaviors, \frangel\ can rapidly discover increasingly-complex behaviors that are likely relevant
to the synthesis task.

Checking program equivalence is undecidable, so \frangel\ approximately characterizes a program's semantics by observing
results on a set of test cases (namely, the I/O examples): a program passing a subset of tests is assumed to ``contain''
the behavior described by those tests. Additionally, if at least one test is passed, then the behavior is assumed to be
relevant to the synthesis task. In this way, every nonempty subset of test cases describes some relevant behavior, and a
program has (at least) that behavior if it passes (at least) those test cases.

At its core, \frangel\ is an adaptive search that randomly samples programs and evaluates them on the I/O examples. It
``remembers'' partially-successful programs, specifically the simplest implementation found so far that has each
relevant behavior. Then, two techniques are used to guide and factorize the search for programs with new behaviors,
respectively:

\begin{enumerate}

    \item \frangel\ \emph{mines code fragments} to learn from the remembered programs. These code fragments can be used
        in future random programs, effectively biasing the search toward the known relevant behaviors. Fragments can be
        combined and modified to produce new behaviors. We also identify and discuss a common property of programs
        called \emph{special-case similarity}, hypothesizing that code fragments from partially-successful programs
        often appear in the desired program, which can justify why this learning approach works well in practice.

    \item \frangel\ also uses \emph{angelic conditions} to decompose the synthesis process into two steps: first find a
        promising program structure, and then determine how to correctly direct its control flow. \frangel\ initially
        generates programs with unspecified control structure conditions (``angelic conditions,'' based on ideas in
        \citet{angelic}) and optimistically evaluates these programs by trying many different control flows. If
        \frangel\ finds an ``angelic program'' with control flows that lead to passing many test cases, it then attempts
        to resolve, or ``fill in,'' the angelic conditions, resulting in a concrete program. This strategy helps
        \frangel\ combine known behaviors with control structures to discover more complex behaviors.

\end{enumerate}

\frangel\ synthesizes Java functions given a collection of libraries, the desired function signature, and a set
of input-output examples. We created a benchmark suite including 90 tasks of widely varying difficulty, using a variety
of libraries and control structure patterns. We also include the 30 tasks used to evaluate \sypet~\cite{sypet}, the
previous approach most similar to ours. While \sypet\ solves 28\% of the 120 tasks within 30 minutes, \frangel\ solves
94\% within 30 minutes and 47\% within 10 seconds. \frangel\ scales to hundreds (sometimes thousands) of components and
can solve tasks requiring up to three control structures with different nesting patterns.

\paragraph{Contributions.}

\begin{itemize}
    \item We present \frangel,\footnote{The code, benchmarks, and results are available at
        \url{https://www.github.com/kensens/FrAngel}~\cite{frangel-url}.} a system for domain-agnostic and
        component-based synthesis from examples (Section~\ref{sec:alg}).
    \item We present a method to \emph{mine code fragments} that are likely to be useful for a given task
        (Section~\ref{sec:fragments}). We also discuss the \emph{special-case similarity} property of programs, and
        argue why this property leads to the effective mining of fragments (Section~\ref{subsec:scs}).
    \item We describe \emph{angelic conditions}, which factorize the synthesis process (Section~\ref{sec:angelic}).
        Programs with angelic conditions are evaluated optimistically with \emph{angelic execution}
        (Section~\ref{subsec:angelicExecution}).
    \item We demonstrate that \frangel\ can generate interesting programs using multiple control structures along with
        libraries containing hundreds of components within seconds (Section~\ref{sec:experiments}).
\end{itemize}

\section{Motivating Example}
\label{sec:motivating}

Suppose we wish to synthesize the target program \code{getRange} (Figure~\ref{fig:getRange}),\footnote{The loop body
could simply be \code{list.add(start + i);}, but \frangel\ currently does not handle autoboxing or unboxing.} which
returns a list of integers from the first argument (inclusive) to the second argument (exclusive).
Figure~\ref{fig:getRange} also lists the test cases (examples), e.g., for the inputs \code{(10, 12)}, \code{getRange}
should return the list \code{[10, 11]}.

One might try to run randomly-generated programs on the test cases, but this strategy alone is infeasible---generating
the entire target program by chance is incredibly unlikely. In fact, for this task, our synthesizer \frangel\ considers
a search space of roughly 30K programs of size 5 (nodes in the abstract syntax tree, or AST), 800K programs of size 6,
and over 10M of size 7. Figure~\ref{fig:getRange} is actually \frangel's solution, which has size 14. In our
experiments, a na\"ive random search was never able to solve this task.

\begin{figure}
    \centering
    \begin{lstlisting}
// #1. (10, 9)  -> []      #4. (10, 12) -> [10, 11]
// #2. (10, 10) -> []      #5. (-2, 2)  -> [-2, -1, 0, 1]
// #3. (10, 11) -> [10]
List<Integer> getRange(int start, int end) {
    ArrayList<Integer> list = new ArrayList<Integer>();
    for (int i = 0; start + i < end; i++)
        list.add(Integer.valueOf(start + i));
    return list;
}   \end{lstlisting}
    \caption{An example target program, \code{getRange}, in Java, with five test cases (examples) in comments.}
    \label{fig:getRange}
\end{figure}

\begin{figure}
    \vspace{-12pt}
    \centering
    \begin{lstlisting}
// Works for tests #1 and #2
List<Integer> getRange(int start, int end) {
    return new ArrayList<Integer>();
}

// Works for test #3
List<Integer> getRange(int start, int end) {
    ArrayList<Integer> list = new ArrayList<Integer>();
    list.add(Integer.valueOf(start));
    return list;
}

// Works for test #4
List<Integer> getRange(int start, int end) {
    ArrayList<Integer> list = new ArrayList<Integer>();
    list.add(Integer.valueOf(start));
    list.add(Integer.valueOf(start + 1));
    return list;
}   \end{lstlisting}
    \caption{Programs that pass some test cases. Our algorithm extracts code fragments from such programs.}
    \label{fig:partialSuccess}
\end{figure}

\paragraph{Mining fragments.} However, it \emph{is} feasible to randomly generate simpler programs that produce the
correct result for some of the test cases---Figure~\ref{fig:partialSuccess} shows a few examples. For instance,
returning an empty list will work for all test cases where \code{start >= end} (e.g., the first and second tests in
Figure~\ref{fig:getRange}). Can we learn from these partial successes to find other relevant behaviors? Indeed, notice
that these simple programs contain important operations used by the target program, such as ${\code{new
ArrayList<Integer>()}}$ and \code{list.add($\blackcirc$)}. This is no coincidence; in many situations, simple programs
that pass some test cases will be similar to the target program. We call this phenomenon \emph{special-case similarity},
discussed in depth in Section~\ref{subsec:scs}. This motivates the idea of \emph{mining fragments}: we extract code
fragments from programs that pass some of the test cases. Newly-generated programs can then use the extracted fragments
or modifications of them. For instance, one fragment extracted from the second program in
Figure~\ref{fig:partialSuccess} is \code{list.add(Integer.valueOf(start))}. This fragment helps generate the third
program, which uses the fragment twice (once with a modification).

Note that the partially-successful programs depend on the specific examples provided. If the test cases only contained
ranges with three or more elements, then the programs in Figure~\ref{fig:partialSuccess} would not pass any test cases.
Our approach is most effective if the examples include a variety of small cases or edge cases, as in
Figure~\ref{fig:getRange}. This requirement is quite natural and is similar to what one might expect from
well-constructed unit tests.

Although \code{list.add(Integer.valueOf(start))} is a useful extracted fragment, generating the full \code{getRange}
program is still difficult. We would have to put the correct modification of the fragment in a loop (changing
\code{start} to \code{start + i}) and simultaneously generate a correct loop condition.

\paragraph{Angelic conditions.} This leads to our second main idea: instead of relying on control structure bodies and
conditions to be correct simultaneously, we decompose the problem by allowing these events to occur in sequence. More
specifically, we randomly generate programs with unspecified control structure conditions and then attempt to fill in
the conditions if the program sketch seems promising. As a concrete example, the following program sketch is an
\emph{angelic program} because it uses an \emph{angelic condition}, denoted \code{\angelic}, in the loop:
\begin{lstlisting}[mathescape=true, frame=none]
List<Integer> getRange(int start, int end) {
    ArrayList<Integer> list = new ArrayList<Integer>();
    for (int i = 0; $\angelic$; i++)
        list.add(Integer.valueOf(start + i));
    return list;
}
\end{lstlisting}

To execute this angelic program, we allow the evaluation of the angelic condition to try different control flows
(numbers of loop iterations) for each test case. For example, this program passes test \#4 (i.e., \code{getRange(10,
12)} returns \code{[10, 11]}) by using 2 loop iterations, and in fact, each test case can be passed using an appropriate
control flow. This observation hints that the angelic program could be a high-level description of the desired behavior.

We now \emph{resolve} the angelic condition by replacing it with a Boolean expression that produces the desired control
flow for all test cases. In this case the expression \code{start + i < end} successfully resolves the angelic condition,
completing the synthesis task.

Despite the enormous search space considered, \frangel\ is able to solve the \code{getRange} task in only 3.5 seconds on
average by mining fragments and using angelic conditions.

\section{The \frangel\ Algorithm}
\label{sec:alg}

\begin{wrapfigure}[16]{r}{0.5\textwidth}
    \centering
    \vspace{-16pt}
    \includegraphics[width=0.49\textwidth]{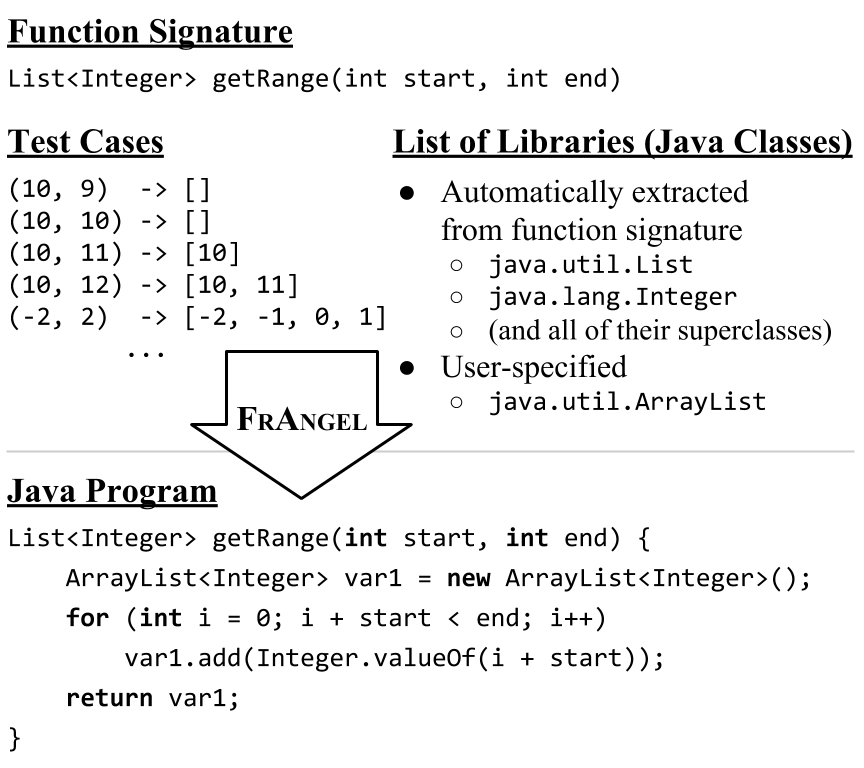}
    \caption{An overview of the problem setting.}
    \label{fig:setting}
\end{wrapfigure}

We now describe the \frangel\ algorithm. We formalize the setting below, summarize the algorithm in
Section~\ref{subsec:structure}, and leave specific details to be discussed in Sections~\ref{subsec:generation},
\ref{subsec:simplification}, \ref{sec:fragments}, and \ref{sec:angelic}.

Figure~\ref{fig:setting} summarizes our problem setting. We focus on synthesizing Java programs, given the signature of
the function to synthesize (the \emph{target program}), a set of test cases (synonymous with \emph{examples} in this
paper), and a list of libraries (Java classes) from which we draw components (i.e., the classes' public methods,
constructors, instance variables, and constants). Our goal is to find a program that is consistent with all examples.
Since we only check the provided examples, the resulting program is not guaranteed to match the user's intent; we assume
that the user will perform code review and/or further testing to verify correctness.

In this paper we use \emph{program} to refer to any function implementation with the desired function signature. A
\emph{test case} $(X,Y)$ contains all inputs $X$ required for a single invocation of the target program, along with the
corresponding outputs $Y$. If the target program changes some mutable inputs, then their new (altered) values are
considered outputs. We say that a program $P$ \emph{passes} a test case $(X,Y)$ if $P$ produces the outputs $Y$ when
invoked with the inputs $X$.

Our implementation makes some simplifying assumptions: we ignore side effects (e.g., filesystem modifications), only
using object equality or user-defined checks to compare program semantics, and we disallow test cases where the desired
action is to throw an exception, although the algorithm in principle can handle this case.

\frangel\ infers relevant libraries (Java classes) by automatically using all types in the target program's signature
and any user-specified types, plus all of their supertypes and nested classes.

\subsection{Algorithm Summary}
\label{subsec:structure}

\begin{algorithm}
    \caption{\frangel}
    \label{alg:structure}
    \begin{algorithmic}[1]
        \Input Target program signature $S$, set of libraries $L$, set of test cases $C$
        \Output A program $P$ that passes all tests in $C$
        \Procedure{\frangel}{$S,L,C$}
            \State $R \gets \emptyset$ \Comment{A set of remembered programs}
            \State $F \gets \emptyset$ \Comment{A set of fragments}
            \RepeatUntilTimeout
                \State $A \gets $ \Call{RandBoolean}{$ $} \Comment{Whether to generate angelic conditions}
                \State $P \gets $ \Call{GenRandomProgram}{$S,L,F,A$} \Comment{Step (1): randomly generate program}
                \State $T \gets $ \Call{GetPassedTests}{$P,C$}
                \If{$T = \emptyset$}
                    \State \textbf{continue}
                \EndIf
                \If{$P$ contains angelic conditions} \Comment{Step (2): resolve angelic conditions}
                    \State $P \gets $ \Call{ResolveAngelic}{$P,L,F,T,C$}
                    \If{failed to resolve conditions}
                        \State \textbf{continue}
                    \EndIf
                    \State $T \gets $ \Call{GetPassedTests}{$P,C$}
                \EndIf
                \State $P \gets $ \Call{SimplifyQuick}{$P,T$}
                \State $T \gets $ \Call{GetPassedTests}{$P,C$}
                \If{$P$ is the simplest program to pass $T$} \Comment{Step (3): mine fragments}
                    \State $R \gets $ \Call{RememberProgram}{$P,R$} \Comment{Add $P$ to $R$ and remove worse programs}
                    \State $F \gets $ \Call{MineFragments}{$R$}
                \EndIf
                \If{$T = C$}
                    \State \textbf{return} \Call{SimplifySlow}{$P,C$} \Comment{Solution found}
                \EndIf
            \EndRepeat
            \State \textbf{return} Failure \Comment{Timeout reached}
        \EndProcedure
    \end{algorithmic}
\end{algorithm}

The overall structure of the \frangel\ algorithm is presented in Algorithm~\ref{alg:structure} and follows the intuition
provided in Section~\ref{sec:motivating}. The core algorithm structure is quite simple, looping over three basic steps
until a solution (i.e., a program passing all test cases) is found or a timeout is reached:
\begin{enumerate}
    \item Randomly generate a program $P$ using previously-mined fragments and angelic conditions.
    \item Resolve angelic conditions in $P$ if any, resulting in a program without angelic conditions.
    \item Mine fragments from the program if it is the simplest program so far to pass some subset of test cases. More
        precisely, for each nonempty subset of test cases, extract fragments from the simplest program so far that
        passes that subset of cases.
\end{enumerate}

\frangel\ also focuses on programs that show signs of (partial) success: \frangel\ only proceeds to step (2) if $P$
passes at least one test case, and step (3) is contingent on successfully resolving all angelic conditions in step (2).
If these conditions are not met, then \frangel\ immediately discards the program and returns to step (1), repeating the
process with a new random program.

Note that \frangel\ only generates angelic conditions for half of the programs. By generating non-angelic control
structures, \frangel\ can extract control structures as fragments without needing to resolve angelic conditions, which
might not succeed for partially-correct programs.

Section~\ref{subsec:generation} describes the random generation of programs in step (1), Section~\ref{sec:angelic}
describes angelic conditions and \textsc{ResolveAngelic} from step (2), and Section~\ref{sec:fragments} discusses
\textsc{RememberProgram} and \textsc{MineFragments} from step (3). Before mining fragments or returning a final program,
\frangel\ uses a simplification procedure to remove unnecessary or overly-verbose code, presented in
Section~\ref{subsec:simplification}.

\subsection{Random Program Generation}
\label{subsec:generation}

A program is a function with the desired signature consisting of some local variable declarations\footnote{The local
variable's type is set to the program's return type half of the time and is chosen randomly among all available types
otherwise. Local variables are initialized with appropriate constants, zero-argument constructors if available, or
\code{null}.} followed by zero or more statements and, if applicable, a return statement with an appropriately-typed
expression. Our grammar for statements and expressions is shown in Figure~\ref{fig:grammar} and specifies a reasonably
large subset of Java. Beyond the restrictions in the grammar, we adhere to (a simplification of) Java's type rules and
explicitly exclude methods such as \code{Object.wait()} and \code{Math.random()} that can stall execution, cause
undesirable side-effects, or lead to nondeterministic programs.

\begin{figure}
    \vspace{-10pt}
    \begin{alignat*}{3}
        & v \enskip && \Coloneqq \enskip \emph{any variable name in scope} \\
        & c  && \Coloneqq \enskip \code{-1} \mid \code{0} \mid \code{1} \mid \code{2} \mid \code{0.0} \mid \code{1.0} \mid
                \code{2.0} \mid \code{true} \mid
                \code{false} \mid \code{null} \mid \code{""} \mid           
                                              \emph{any user-provided constant} \\
        & op && \Coloneqq \enskip \code{+} \mid \code{-} \mid \code{*} \mid \code{/} \mid \code{\%} \mid
                \code{\&\&} \mid \code{||} \mid \code{==} \mid \code{<} \mid \code{<=} \\
        & f  && \Coloneqq \enskip \code{(}e\code{).}\emph{method}\code{(}e^*\code{)} \mid \emph{ClassName}\code{.}\emph{staticmethod}\code{(}e^*\code{)} \\
        & e  && \Coloneqq \enskip v \mid c \mid \code{(}e\code{) }op\code{ (}e\code{)} \mid
                \code{!(}e\code{)} \mid f \mid \code{(}e\code{)[}e\code{]} \mid \code{(}e\code{).}\emph{field} \mid \\[-0.4em]
            &&& \phantom{{}\Coloneqq{}} \enskip \emph{ClassName}\code{.}\emph{field} \mid
                    \code{new }\emph{ClassName}\code{(}e^*\code{)} \\
        & s  && \Coloneqq \enskip v\code{ = }e\code{;} \mid \code{(}e\code{)[}e\code{] = }e\code{;} \mid f\code{;} \mid
                    \code{if (}e\code{) \string{ $\color{black}s^+$ \string}} \mid \\[-0.4em]
            &&& \phantom{{}\Coloneqq{}} \enskip \code{for (int i = 0; }e\code{; i++) \string{ $\color{black}s^+$ \string}} \mid
                    \code{for (}\emph{ClassName}\code{ }v\code{ : }v\code{) \string{ $\color{black}s^+$ \string}}
    \end{alignat*}
    \caption{\frangel\ grammar for expressions $e$ and statements $s$. We use $e^*$ to represent zero or more
    comma-separated expressions, and $s^+$ for one or more statements.}
    \label{fig:grammar}
\end{figure}

Recall that \frangel\ collects a set of previously-mined fragments (produced by \textsc{MineFragments}, detailed in
Section~\ref{sec:fragments}) that can be used as building blocks, possibly with modifications, when generating new
programs. A \emph{fragment} is a complete subtree of a previous program's abstract syntax tree (AST), e.g., the
code \code{return a + b.foo(c);} contains fragments including \code{b.foo(c)} and \code{c}.

We now describe how a program's statements and expressions are randomly generated, first discussing a simpler case
without fragments, and then the case with fragments.

\paragraph{Procedure A (Basic Generation).} To generate an expression or statement, we first choose a kind of expression
(variable, constant, operator, or function call) or statement (assignment, function call, \code{if} statement, or
\code{for} loop) uniformly at random. Further decisions, such as the specific operator or function to use, are made
uniformly at random among options in the grammar that typecheck. Subexpressions and substatements are generated
recursively.

This procedure also takes a \emph{size} parameter that upper-bounds the size (number of AST nodes) of the resulting AST
or subtree. The tree's root node must be chosen to adhere to the size constraint, taking into account the number of
children it would require. Then, the node's remaining size is randomly partitioned among its children. The \emph{size}
parameter is primarily used to avoid generating infinitely-large trees, since otherwise this random generation procedure
is a Galton-Watson process~\cite{galtonwatson}. In our experiments, programs are generated with size at most 40. This is
a much larger limit than necessary, as most of our benchmarks are solved by programs smaller than size 20.

\paragraph{Procedure B (Generation with Fragments).} We now explain how \frangel\ takes into account the mined code
fragments. Empirically, correct programs often contain mined fragments (perhaps with different variable names) as exact
copies or with modifications (we elaborate on this observation in Section~\ref{sec:fragments}). However, correct
programs can also include code that is not similar to any mined fragment. To handle these cases, \frangel\ can use an
existing fragment as-is or with random parts modified, or generate code from scratch. This strategy biases the
random program generation toward the mined fragments while maintaining the flexibility of random search.

Specifically, \frangel\ uses the following procedure to generate an expression or statement $X$:

\begin{enumerate}

    \item With $\frac{1}{2}$ probability, we generate the root of $X$ randomly from scratch using \emph{Procedure A},
        except that subtrees are generated recursively with \emph{Procedure B}.

    \item Otherwise, we sample uniformly, from the set of mined fragments, an expression or statement $X'$ of the same
        type as $X$ (defaulting to the previous case if no such $X'$ exists). Then,

        \begin{enumerate}

            \item We make the variable names in $X'$ compatible with the surrounding program $P$ by changing the names
                in $X'$ to match existing variable names in $P$ of the same type. If such names do not exist, we
                accommodate by first declaring new variables in $P$ for the missing types.

            \item With $\frac{1}{2}$ probability, we use the entire fragment $X'$ as-is by setting $X=X'$.

            \item Otherwise, we generate $X$ to be a random modification of $X'$. More precisely, a rooted connected
                component of $X$ will match $X'$, while the rest is generated randomly without fragments. We start by
                making the root of $X$ the same as the root of $X'$; then, as we walk down the tree, at each node we
                continue to make $X$ match $X'$ with probability $\frac{3}{4}$, and otherwise we generate that subtree
                of $X$ randomly using \emph{Procedure A}.

        \end{enumerate}
\end{enumerate}

\subsection{Program Simplification}
\label{subsec:simplification}

\frangel\ applies two simplification strategies (\textsc{SimplifyQuick} and \textsc{SimplifySlow} in
Algorithm~\ref{alg:structure}) to the mined fragments and the final solution. This eliminates unnecessary code,
resulting in a slight speedup (by increasing the relevancy of mined fragments) and more natural final solutions.

Before mining fragments from a program $P$, \frangel\ simplifies $P$ with \textsc{SimplifyQuick}. Let $T$ be the set of
test cases originally passed by $P$, and for any AST node $N$, let $N_\top$ represent the AST subtree rooted at $N$.
\textsc{SimplifyQuick} can be described as follows:

\begin{enumerate}
    \item For every node $N$ in the AST of $P$:
        \begin{enumerate}
            \item Consider all of the following AST subtrees, which will act as ``replacements'' for $N_\top$:
                \begin{itemize}
                    \item The empty tree, if $N_\top$ is a statement in a block.
                    \item Single-node trees corresponding to all variables and constants in the grammar.
                    \item $D_\top$ for all descendant nodes $D$ of $N$.
                \end{itemize}
            \item For each of the above ``replacement'' subtrees $R$:
                \begin{enumerate}
                    \item Temporarily replace $N_\top$ with $R$, if doing so follows Java type and syntax rules, and if
                        $R$ is smaller than $N_\top$ (comparing first by the number of AST nodes and then by code
                        length).
                    \item If the new code passes all tests in $T$, then make the replacement permanent. Otherwise, undo
                        the temporary replacement.
                \end{enumerate}
        \end{enumerate}
    \item Repeat until $P$ cannot be simplified further (i.e., all remaining replacements are unsuccessful).
\end{enumerate}

Before returning a final program, \frangel\ uses a more extensive simplification strategy, \textsc{SimplifySlow}. This
is the same as \textsc{SimplifyQuick}, except with the following modifications:

\begin{itemize}
    \item Only statements, control structure conditions, and the return expression of $P$ can be replaced. Hence,
        \textsc{SimplifySlow} focuses on larger-scale changes compared to \textsc{SimplifyQuick}.
    \item Replacement subtrees $R$ are generated randomly.
    \item Instead of proceeding until no more replacements are possible, \textsc{SimplifySlow} runs for 10\% of the
        total time spent before \textsc{SimplifySlow}.
    \item After the time limit is reached, \textsc{SimplifySlow} concludes with one run of \textsc{SimplifyQuick}.
\end{itemize}

Although the final program behaves identically on the test cases before and after \textsc{SimplifySlow}, simplification
can greatly improve readability, e.g., by replacing \code{(str + "").equalsIgnoreCase(str)} with the equivalent
\code{str != null}. \textsc{SimplifySlow} can also improve awkward expressions that might not generalize fully. For
instance, if \code{n} is an \code{int}, then replacing \code{1 < n + 1} with \code{0 < n} can eliminate an integer
overflow bug that might not be evident from executions on the given test cases. In this way, simplification can reduce
the number of false positives produced by \frangel\ (i.e., programs that pass all test cases but do not completely match
the user's intent).

\section{Mining Fragments}
\label{sec:fragments}

As we observed in the motivating example (Section~\ref{sec:motivating}), programs passing some of the tests often share
code with the target program. \frangel\ uses this observation by mining fragments from such programs and using those
fragments when generating new programs. This biases the search toward previous partial successes and potentially new
relevant behaviors, such as the target program.

Given a program $P$, a \emph{fragment} is any complete subtree of $P$'s abstract syntax tree (AST). For example, in the
program \code{return a + b.foo(c)}, the fragments are \code{a + b.foo(c)}, \code{b.foo(c)}, \code{a}, \code{b}, and
\code{c}. Neither \code{+} nor \code{b.foo($\blackcirc$)} are fragments because they are not \emph{complete} subtrees.
However, when \frangel\ uses fragments to generate new programs using \emph{Procedure B} as described in
Section~\ref{subsec:generation}, it will be able to generate code like \code{b.foo($\blackcirc$)}, where $\circ$ is
generated randomly from scratch.

In Algorithm~\ref{alg:structure}, \textsc{RememberProgram} chooses which programs to ``remember,'' and
\textsc{MineFragments} extracts all fragments from all currently-remembered programs. A program is remembered if it is
the simplest program encountered by \frangel\ so far that passes some nonempty subset of test cases.\footnote{The
program can also pass other test cases, so a program can be the simplest for many subsets of tests simultaneously. In
other words, a program is not remembered if there is a simpler program passing a superset of tests.} Simplicity is
measured by the number of AST nodes and code length. Note that a program that is remembered at one point can later be
``forgotten'' when \frangel\ finds an even-simpler program passing at least the same test cases. Hence, the set of
remembered programs gradually improves as \frangel\ runs---it grows in size and captures more behaviors as more distinct
subsets of test cases are passed,\footnote{Although there are exponentially many subsets of test cases, empirically the
number of remembered programs is less than 100, even for tasks with 12 test cases.} and the remembered programs can be
replaced if better (simpler) versions are found. Also note that \textsc{MineFragments} returns a \emph{set} of
fragments, i.e., a fragment that appears multiple times in the remembered programs occurs only once in the set of mined
fragments.

While the provided libraries might have hundreds of components, mining fragments allows \frangel\ to focus on the ones
that are likely to be useful for the current synthesis task. Furthermore, since \frangel\ can mine fragments from code
that was generated using previously-mined fragments, the fragments often grow larger by composition and better through
random improvements.

\subsection{Special-Case Similarity}
\label{subsec:scs}

\begin{wrapfigure}[8]{r}{0.5\textwidth}
    \centering
    \vspace{-13pt}
    \includegraphics[width=0.5\textwidth]{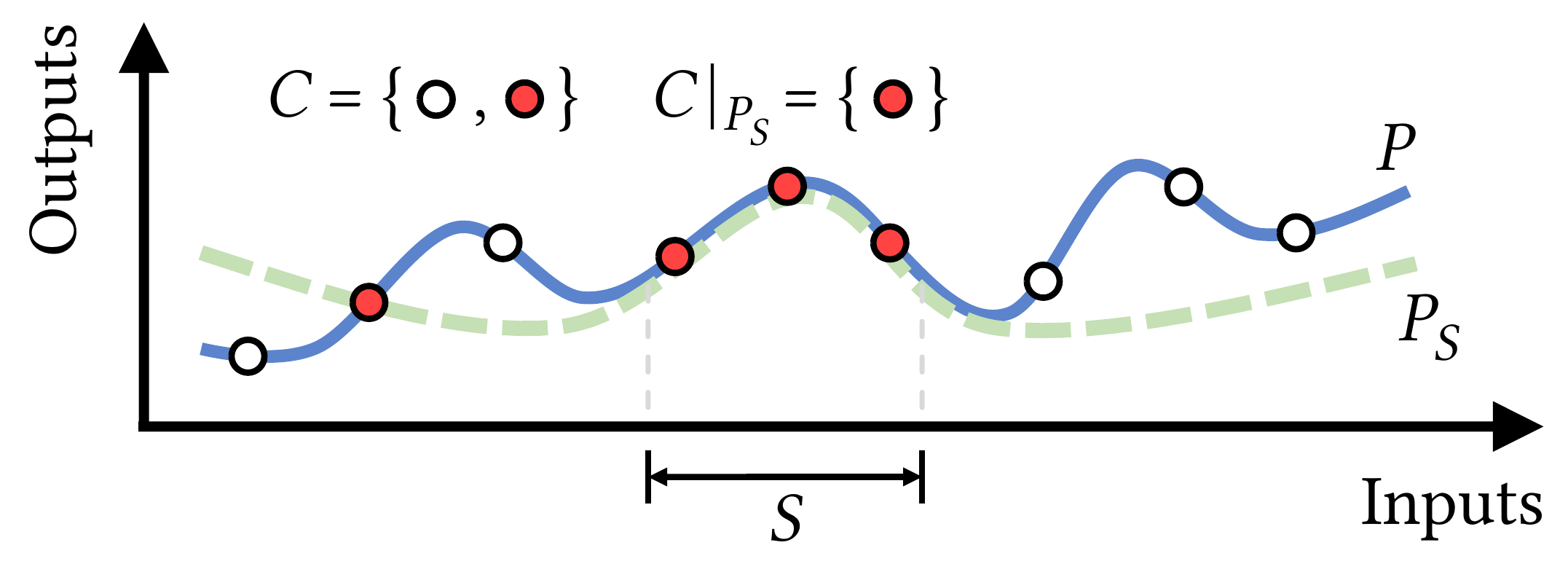}
    \caption{Diagram illustrating \emph{special-case similarity} and \emph{exposure} using an analogy on the real line.}
    \label{fig:scs}
\end{wrapfigure}

Why should we expect the mined fragments to be useful for synthesizing the solution? To answer this, we describe a
common property of programs called \emph{special-case similarity} (SCS).

We begin with an analogy on the real line for intuition. In Figure~\ref{fig:scs}, the target program $P$ maps inputs to
outputs, represented as a function (solid blue curve). Consider a set $S$ of inputs. Let $P_S$ (dashed green curve) be
the ``simplest'' function matching $P$ for the inputs in $S$. If $P_S$ is ``similar'' to $P$, then $P$ is special-case
similar~(SCS) with respect to $S$, and we say that $S$ is a \emph{special case} with the corresponding
\emph{special-case program} $P_S$. That is, $P$ is SCS if there are inputs (e.g., $S$) where the simplest solution
(i.e., $P_S$) is similar to $P$.

In the realm of programs, we measure \emph{simplicity} first by the number of AST nodes and then by the code length, but
other reasonable metrics exist, e.g., preferring natural code as done by \citet{nsc}. We say that two programs are
\emph{similar} if there is nontrivial overlap, i.e., duplicate connected subgraphs of the AST, between their sets of
fragments. For instance, the fragment \mbox{\code{a + b.foo(1)}} significantly overlaps with \code{bar(1 + b.foo(c))}
since the 3-node portion \mbox{\code{$\blackcirc$ + b.foo($\blackcirc$)}} is equal. However, it only has trivial overlap
with \code{b.baz()} since only node \code{b} is duplicated.

Ideally, we want \frangel\ to remember and mine fragments from special-case programs such as $P_S$. Compared to the
target program $P$, special-case programs are easier to find because they are simpler and more likely to be generated.
By mining fragments from remembered special-case programs, which by definition are similar to $P$, \frangel\ guides its
search toward $P$ even without explicit knowledge of $P$, thus reaching a final solution much faster.

The following definition formalizes this intuition of SCS.

\begin{defn}[special-case similarity (SCS)]

    Assume a fixed grammar\, $G$ over programs. Consider a program $P\in G$. For any nonempty set $X$ of inputs to $P$,
    let $P_X\in G$ be the simplest program such that $P$ and $P_X$ have identical functionality when restricted to $X$.
    We define $P$ to be \emph{special-case similar with respect to} a nonempty set $S$ of inputs (a ``special case'') if
    $P_S$ (a ``special-case program'') is similar to $P$ and $P_S \ne P$. More generally, we say that $P$ is
    \emph{special-case similar (SCS)} if there exists such a set $S$.

\end{defn}

Note that if $P$ is SCS, then a special-case program $P_S$ must be strictly simpler than $P$, and in most cases, $P_S$
is actually a \emph{simplification} of $P$. Also, $P$ can be SCS with respect to multiple sets of inputs.

Our specific notion of similarity is based on the way that \frangel\ makes use of fragments. As a general intuition,
$P_S$ is ``similar enough'' to $P$ if the knowledge of $P_S$ allows the synthesis algorithm to produce $P$ faster or
with higher probability. We also point out that the \frangel\ algorithm itself does not use any notion of similarity,
which is discussed here only for intuitive justification for fragment mining.

\paragraph{Exposure}

Whether \frangel\ remembers a specific special-case program $P_S$ depends largely on the user-provided test cases $C$,
denoted by circles with any fill color in Figure~\ref{fig:scs}. Recall that \frangel\ remembers the simplest program
that passes each nonempty subset of test cases. Hence, as long as $P_S$ is the simplest program solving some subset
$C'\subseteq C$ with $C'\ne\emptyset$, then \frangel\ is guaranteed to remember (and never forget) $P_S$ once it is
randomly generated as a candidate program. In this case, we say that $S$ (i.e., the special case corresponding to $P_S$)
is \emph{exposed} by the test cases $C$.

It actually suffices to only consider the test cases in $C$ that are consistent with $P_S$, denoted by circles with red
fill in Figure~\ref{fig:scs}. Let the notation $C|_{P_S}$ refer to that subset of $C$. Then, $C$ exposes $S$ if and only
if $C|_{P_S} \ne \emptyset$ and $P_S$ is the simplest program solving $C|_{P_S}$. To see this, note that if $P_S$ is the
simplest program for some nonempty $C'\subseteq C$, then all elements of $C'$ are in $C$ and are consistent with $P_S$,
i.e., $C'\subseteq C|_{P_S}$. Hence, any program solving $C|_{P_S}$ must also solve $C'$. Since $P_S$ is the simplest
program for $C'$, the simplest program for $C|_{P_S}$ cannot be simpler than $P_S$. Therefore, as $P_S$ is consistent
with $C|_{P_S}$ by construction, $P_S$ is also the simplest program for $C|_{P_S}$. This leads to the following
definition:

\begin{defn}[exposure]

    Let $P$ be a program that is SCS with respect to a special case $S$ (a set of possible inputs). Let $C$ be a set of
    test cases that are consistent with $P$, and let $C|_{P_S}$ be the subset of test cases in $C$ that are also
    consistent with $P_S$ (the special-case program for $S$). We say that $C$ \emph{exposes} the special case $S$ if
    $C|_{P_S}\ne\emptyset$ and $P_S$ is the simplest program consistent with $C|_{P_S}$.

\end{defn}

\paragraph{Examples}

The \code{getRange} program (Figure~\ref{fig:getRange}) is SCS because the programs in Figure~\ref{fig:partialSuccess}
are all special-case programs. For instance, the second program (duplicated below) is the simplest program for the
special case where \code{end} equals \code{start + 1}. This is similar to the target program because both contain
fragments of the form \code{list.add(Integer.valueOf($\blackcirc$))}:

\begin{lstlisting}
// Works for test case #3, and any other input where end == start + 1
List<Integer> getRange(int start, int end) {
    ArrayList<Integer> list = new ArrayList<Integer>();
    list.add(Integer.valueOf(start));
    return list;
}
\end{lstlisting}

This special case is also exposed by the test cases in Figure~\ref{fig:getRange}, since the above program is also the
simplest in the grammar to pass test case \#3. If test case \#3 were removed, then the special case would no longer be
exposed. Overall, special cases are more likely to be exposed if the test cases cover a variety of scenarios, and adding
more tests cannot decrease exposure.

As another example, consider a \code{capitalize} program that accepts a String \code{str} and returns
\code{str.toUpperCase().charAt(0) + str.toLowerCase().substring(1)} if \code{str} is nonempty or \code{""} otherwise.
This has several special-case programs, some of which are shown below with the corresponding special cases in comments:
\begin{lstlisting}
return ""; // str is empty

return str.toUpperCase().charAt(0) + str.toLowerCase().substring(1); // str is nonempty

return str.charAt(0) + str.toLowerCase().substring(1); // First char needs no change

return str.toLowerCase(); // First character is uncased

return str.toUpperCase(); // All characters after the first are uncased
\end{lstlisting}
Note that fragments from the simpler special-case programs are useful in generating the larger special-case programs,
which in turn help \frangel\ generate the target program. By using fragments, \frangel\ can quickly find programs with
new complex behaviors that are relevant to the task.

\paragraph{Relation to mining fragments.}
\newcommand{\prop}[1]{\textbf{(P{#1})}}

Special-case similarity justifies our approach of mining fragments by the following informal argument. When \frangel\
remembers a program $P'$ that passes a subset of test cases $T\subseteq C$, we optimistically assume three properties:
\begin{itemize}[leftmargin=3em]
    \item[\prop1] The target program $P$ is SCS with a special case $S$ exposed by the test cases $C$
    \item[\prop2] The subset $T$ is responsible for exposing $S$, i.e., $T=C|_{P_S}$ and $P_T=P_S$
    \item[\prop3] $P' = P_T$, i.e., the program remembered is actually the simplest solution for $T$
\end{itemize}

If these properties hold, then $P' = P_T = P_S$ is similar to the target program $P$, so fragments from $P'$ are useful
for generating $P$. Property \prop1 is determined by the task (which might not be SCS) and the user's choice of examples
(which might not expose special cases, if any); \frangel\ has little control over these. But if \prop1 holds, then
\prop2 must hold for some $T\subseteq C$. Finally, as soon as \frangel\ randomly generates $P_T$ as a candidate program,
\prop3 is guaranteed.

Of course, the three properties do not always hold for every remembered program, so some mined fragments might not help
\frangel\ find the target program. In practice however, the positive effect of the helpful fragments generally outweighs
the negative effect of the unhelpful ones.

\paragraph{Why SCS is common.} Special-case programs often arise through simplifications in many natural ways (these are
not mutually exclusive, and multiple simplifications can be applied at once):

\begin{itemize}
    \item \emph{Fixpoints}: If a type-preserving operation $f$ is applied to an expression $e$, and $f$ has a fixpoint,
        then we can omit $f$ for all inputs that cause $e$ to be a fixpoint. For example, we can replace
        \code{str.toLowerCase()} with \code{str} whenever \code{str} contains no uppercase characters.

    \item \emph{Simpler expressions}: If an expression $e$ is equivalent to a simpler expression $e'$ under certain
        conditions, then whenever those conditions hold, we can replace $e$ with $e'$. For example, we can replace
        ${\code{arr[arr.length - 1]}}$ with \code{arr[2]} whenever \code{arr} has length 3.

      \item \emph{Degenerate values}: Some operations, especially arithmetic ones, can be simplified if an operand is
          degenerate. For example, we can replace \code{a + b} with \code{b}, and \code{a * b} with \code{0}, whenever
          \code{a} is \code{0}.

    \item \emph{Control structures}: Inputs follow a specific control flow through a program's control structures, so we
        can partition inputs based on their control flows. For each group of inputs, we can tailor the program to that
        control flow by removing, simplifying, or unrolling control structures.

    \item \emph{Edge cases}: If the program has edge-case logic, then all such logic can be omitted for general-case
        inputs. Also, the edge cases can sometimes be handled without the general-case code.
\end{itemize}

The above points show that simplifications are often possible, leading to similar and simpler programs. Usually, such
simplifications result in special-case programs and thus special-case similarity. However, occasionally there is an even
simpler program that is completely different. For instance, consider the task of returning the expression ${\code{(a *
b) - a - b}}$. For the special case where \code{a} is \code{1}, we might simplify this to ${\code{(b) - a - b}}$. But
the simplest program for this special case is actually \code{return -1;}, which is not at all similar to the target
program. Hence, this program is not SCS with respect to the special case where \code{a} is \code{1}.

\subsection{Limitations of Mining Fragments}
\label{subsec:limitationsFragments}

Special-case similarity does not always hold. For example, the following program (a benchmark used by
\sypet~\cite{sypet}) is not SCS:
\begin{lstlisting}[frame=none]
Rectangle2D scale(Rectangle2D rect, double x, double y) {
    return new Area(rect).createTransformedArea(
            AffineTransform.getScaleInstance(x, y)).getBounds2D();
}
\end{lstlisting}
Note that this program mostly contains datatype conversions, which often cannot be simplified---they cannot be
fixpoints, and they are already the simplest way to convert between the types. We could eliminate
\code{.createTransformedArea(...)} whenever \code{x} and \code{y} are \code{1.0}, but the program then collapses to
\code{return rect;}, which is not similar to the target program (the overlap is trivial---since reasonable programs
should use their arguments, the fragment \code{rect} provides no useful information).

Furthermore, mining fragments is not always successful, since the following issues might occur:

\begin{itemize}

    \item \emph{Task not SCS}: Without special-case programs, the mined fragments might be completely unrelated to the
        target program. In this scenario, property \prop1, described above, fails.

    \item \emph{Unexposed special cases}: If the test cases are insufficient to expose any special cases, failing
        \prop1, then \frangel\ is unlikely to remember or benefit from special-case programs.

    \item \emph{Too much noise}: Even if \frangel\ remembers a special-case program $P_T$ that is similar to $P$, there
        might be many other subsets of test cases solved by irrelevant programs not similar to $P$. That is, \prop2
        might hold very rarely for subsets of test cases solved by candidate programs. Mining fragments from those
        irrelevant programs can dilute the positive signal provided by $P_T$. This often arises when there are many
        wrong ways of computing the right result, such as in functions that return Booleans or small integers. For those
        tasks, \frangel\ often remembers programs like \code{return 1 + 2;}, but fragments from such programs are
        usually not helpful.

    \item \emph{Special-case programs too complex}: \frangel\ cannot benefit from special-case programs that are too
        complex to generate in the first place using the currently-mined fragments. In this case, \prop3 might not hold
        within a reasonable amount of time, thus stalling progress.

    \item \emph{Special cases eliminating structure}: Consider summing the elements of an array. A good solution stores
        the running total in a local variable, for instance with ${\code{ans += arr[i];}}$ in a loop. However,
        special-case programs eliminate that variable by summing elements directly, e.g., \code{return arr[0] + arr[1] +
        arr[2];}. More generally, despite being helpful, the mined fragments might not reveal all aspects of the target
        program.

\end{itemize}

\section{Angelic Conditions}
\label{sec:angelic}

Recall that \frangel\ uses an adaptive random search over programs. Angelic conditions factorize this search, allowing
\frangel\ to first find a code sketch without control structure conditions (an \emph{angelic program}) and later search
for the correct conditions once a satisfactory angelic program is found. This factorization helps \frangel\ discover
programs with complex behaviors involving control structures. When angelic conditions are used with the strategy of
mining fragments, \frangel\ can also identify the ``purposes'' of certain fragments---for instance, some functionality
is only sometimes necessary and should go inside a conditional, and some functionality is meaningful if repeated and can
go inside a loop. Recognizing these situations allows \frangel\ to generalize and extend known behaviors.

An \emph{angelic condition}, denoted by \code{\angelic}, can appear in an angelic program anywhere a control structure
condition would normally appear. Informally, each time an angelic condition is evaluated, it can choose to be
\code{true} or \code{false}, whichever would lead to a correct result if possible (as if angels were controlling the
program's execution).

To provide intuition, consider the task of summing the positive values in an array, so the input \code{[-1.2, 3.4]}
produces the output \code{3.4}. Suppose we generated the angelic program in Figure~\ref{fig:angelic}, which is a correct
solution but with angelic conditions in the \code{for} loop and \code{if} statement. Hence, it represents summing an
unspecified subset of the array elements. The precise execution of an angelic program can be described by a \emph{code
path}, which lists how the angelic conditions evaluate in the order that they are encountered. For instance, this
angelic program produces the correct output with the code path \emph{TFTTF}: the loop is entered (\emph{T}),
\code{arr[0]} is not summed (\emph{F}), the loop is continued (\emph{T}), \code{arr[1]} is summed (\emph{T}), and the
loop exits (\emph{F}). The existence of a code path that produces the correct output hints that the angelic program
might be correct, i.e., that the angelic conditions can be \emph{resolved} (filled in) to create a correct non-angelic
program. 

\begin{figure}
    \centering
    \begin{lstlisting}[mathescape=true]
double sumPositiveDoubles(double[] arr) {
    double sum = 0.0;
    for (int i = 0; $\angelic$; i++)
        if ($\angelic$)
            sum = sum + arr[i];
    return sum;
}   \end{lstlisting}
    \caption{An example angelic program with two angelic conditions.}
    \label{fig:angelic}
\end{figure}

Using this intuition, \frangel\ optimistically executes angelic programs. In \textsc{GetPassedTests} from
Algorithm~\ref{alg:structure}, if $P$ is angelic, \frangel\ considers $P$ to pass a test case $(X,Y)$ if it finds a code
path that causes $P$ to produce the outputs $Y$ when given the inputs $X$. Note that this is not based on the
\emph{existence} of a good path, only whether \frangel\ finds one.

In Section~\ref{subsec:bitstring} below, we formally describe how angelic programs are executed using code paths. In
Section~\ref{subsec:angelicExecution}, we provide an algorithm that searches over code paths to determine if an
angelic program passes a given test case. Finally, Section~\ref{subsec:resolvingAngelic} discusses how we resolve
angelic conditions.

\subsection{Bitstring Code Paths}
\label{subsec:bitstring}

We previously explained that code paths describe a particular way of executing an angelic program. We now formalize this
idea and introduce some terminology.

Code paths are represented as bitstrings over $\{T, F\}$, representing \code{true} and \code{false}, respectively. When
we execute an angelic program $P$ using a bitstring code path $b_1\dots b_n$, we make the $i$-th angelic condition (as
encountered during program execution) evaluate to $b_i$ for $1\le i\le n$, or \code{false} for $i>n$. With this
convention, we do not need to consider any bitstring that ends in \emph{F}, since it is semantically identical to the
bitstring obtained by removing trailing \emph{F}s, e.g., the bitstrings \emph{TFTFF} and \emph{TFT} lead to identical
executions. We choose this convention because repeatedly choosing the \emph{F} branch of control structures will
eventually lead to the method's termination, while repeatedly choosing the \emph{T} branch can cause infinite loops.

When \frangel\ begins to execute an angelic program $P$ using a bitstring code path, we call this the \emph{attempted}
code path. Then, there is also a corresponding \emph{actual} code path that takes into account the time of $P$'s
termination (returning or crashing). The actual code path will always be either the attempted code path with zero or
more \emph{F}s appended, or a prefix of the attempted code path. For instance, suppose we run the angelic program in
Figure~\ref{fig:angelic} with the input \code{[-1.2, 3.4]} and the attempted code path \emph{TTT}. After the three
\emph{T}s, we need two more \emph{F}s to reach the \code{return} statement: one to bypass the inner \code{if} statement,
and one more to break out of the loop. Hence, the actual code path is \emph{TTTFF}. However, if the input were instead
the empty array \code{[]}, then the actual code path would be \emph{TT} because the program crashes when accessing
\code{arr[0]}, before reaching the third \emph{T}. We will use actual code paths to prune a search, described below.

\subsection{Angelic Execution}
\label{subsec:angelicExecution}

To determine if an angelic program $P$ passes a given test case, \frangel\ searches for a code path that causes $P$ to
behave consistently with that test case. This must be done carefully; a brute-force search over all code paths (e.g.,
via recursive backtracking) is exponential-time if an angelic condition is inside a loop, since the angelic condition
can evaluate to \code{true} or \code{false} independently on each iteration. We instead use an enumerative partial
search over a bounded number of code paths. This prioritizes \emph{simple} code paths, i.e., shorter paths with fewer
\emph{T}s. Focusing on such paths allows us to assess simple test cases with short control flow traces, since following
the \emph{T} branch of a control structure causes more code to be executed than the \emph{F} branch (at least for loops
and \code{if} statements\footnote{Note that our grammar (Figure~\ref{fig:grammar}) does not include \code{else}
statements, although these ideas can be extended to \code{else} and \code{else if} structures. For instance, we could
force the smallest block to execute in the \emph{F} case.}).

Following this intuition, we enumerate bitstrings ordered first by increasing number of \emph{T}s, and then in
lexicographical order where $T<F$, so \emph{TFT} comes before \emph{TFFT}, which comes before \emph{TTT}. As mentioned
in Section~\ref{subsec:bitstring}, we do not include bitstrings that end in \emph{F}. Each enumerated bitstring is
executed as an attempted code path, and the actual code path is recorded. During this enumeration, we also avoid all
bitstrings that start with a previously-recorded actual code path. For instance, if we attempt the code path \emph{T}
and obtain the actual path \emph{TFF}, we do not need to later attempt \emph{TFFT}, since we already know that
termination occurs after the \emph{TFF}. This allows \frangel\ to quickly prune the search and avoid all redundant code
executions. \frangel\ declares that the angelic program $P$ passes the test case as soon as it finds some code path that
causes the test to pass.

After executing simple code paths, further executions give diminishing returns as there are more paths of similar
complexity. Hence, a constant maximum number $M$ of paths are attempted; if none are successful, then \frangel\ declares
that $P$ does not pass the test case. (Because this is only a partial search, there could be a correct but unattempted
code path.) This gives us a parameterized tradeoff: if $M$ is larger, angelic execution takes longer but searches more
complex code paths. We set $M=55$ in our experiments.\footnote{If we have an \code{if} statement inside a \code{for}
loop (and no other control structures), then the 55th enumerated code path is 8 consecutive \emph{T}s. That is, within
55 code paths we will have tried all ways of performing at most 4 loop iterations.}

As a concrete example, consider executing the angelic program in Figure~\ref{fig:angelic} using the input \code{[-1.2,
3.4]}. The process is summarized below:

\newcommand{\myblue}{\color{blue!70!black}} 
\newcommand{\mygreen}{\color{green!40!black}} 
\newcommand{\myred}{\color{red!50!black}} 
\begin{center}
\setlength{\tabcolsep}{8pt}
\begin{tabular} {c l l l}
    \# \emph{T}s & Attempted path & Actual path & Output\\
    0 & $\epsilon$  & \emph{\myblue F}       & \code{\color{black}0.0} \\
    1 & \emph{T}    & \emph{\mygreen TFF}     & \code{\color{black}0.0} \\
      & \multicolumn{3}{l}{\color{gray} Skip \emph{FT}, \emph{FFT}, \emph{FFFT}, etc. (starts with \emph{\myblue F}\,)} \\
    2 & \emph{TT}   & \emph{\myred TTF}     & \code{\color{black}-1.2} \\
      & \emph{TFT}  & \emph{TFTFF}   & \code{\color{black}0.0} \\
      & \multicolumn{3}{l}{\color{gray} Skip \emph{TFFT}, \emph{TFFFT}, etc. (starts with \emph{\mygreen TFF}\,)} \\
      & \multicolumn{3}{l}{\color{gray} Skip \emph{FTT}, \emph{FTFT}, etc. (starts with \emph{\myblue F}\,)} \\
    3 & \emph{TTT}  & \emph{TTTFF}   & \code{\color{black}-1.2} \\
      & \multicolumn{3}{l}{\color{gray} Skip \emph{TTFT}, \emph{TTFFT}, etc. (starts with \emph{\myred TTF}\,)} \\
      & \emph{TFTT} & \emph{TFTTF}   & \code{\color{black}3.4} \\
      & \multicolumn{3}{l}{\color{gray} Correct output; test case \code{\color{gray}[-1.2, 3.4]} passed.} \\ 
\end{tabular}
\end{center}

Note that after attempting \emph{T}, the next smallest bitstring with one \emph{T} is \emph{FT}, which we skip because
it starts with \emph{F} (the actual code path for the previously attempted path $\epsilon$). \frangel\ simultaneously
skips the infinite set of bitstrings \emph{FFT}, \emph{FFFT}, etc., for the same reason. Angelic execution ends after
attempting \emph{TFTT}, since this results in the correct output of \code{3.4} for this test case. We only needed 6
attempts to find a good code path of length 5 due to the heavy pruning of enumerated bitstrings.

So far, we have only described how an angelic program is executed on a single test case. Naturally, we would like to
repeat this process for all test cases, but this can be slow as it involves trying many code paths for each test case.
To avoid excessive slowdown, \frangel\ terminates the overall angelic execution early when a $1-\sigma$ fraction of the
test cases have already failed (we set $\sigma=0.75$ in our experiments). Note that even a correct angelic program might
not pass all test cases if the required code path for some test case is too complex.

If an angelic program does pass at least a $\sigma$-fraction of the test cases, then \frangel\ proceeds to the next
step: resolving the angelic conditions to produce a non-angelic program.

\subsection{Resolving Angelic Conditions}
\label{subsec:resolvingAngelic}

Once we have a promising angelic program $P$, the final step is to resolve the angelic conditions, replacing them with
concrete expressions to obtain a non-angelic program passing the same tests.

Pseudocode is given in Algorithm~\ref{alg:resolve}. To resolve an angelic condition, we replace it with random Boolean
expressions generated using \emph{Procedure B} in Section~\ref{subsec:generation}. We evaluate the resulting program on
the test cases, using angelic execution if appropriate. If it passes all test cases that $P$ previously passed, then we
have successfully resolved a condition and can proceed to resolve the next one. When resolving an angelic condition, we
impose a time limit roughly proportional to the amount of time elapsed since the previous attempt to resolve conditions
in a different angelic program.

We first try to resolve conditions from the innermost to the outermost. If that fails, we try again in the reverse
order. If we are unable to resolve conditions, the program is discarded.

\begin{algorithm}
    \caption{\textsc{ResolveAngelic}}
    \label{alg:resolve}
    \begin{algorithmic}[1]
        \Input Angelic program $P$, set of libraries $L$, set of fragments $F$, initially-passed tests $T$, all tests
        $C$
        \Output A non-angelic program that passes all of $T$, or Failure
        \Procedure{ResolveAngelic}{$P,L,F,T,C$}
            \While{$P$ contains angelic conditions}
                \State \emph{success} $\gets \bot$
                \RepeatUntilTimeout
                    \State $B \gets $ \Call{GenBooleanExpression}{$L, F$}
                    \State $P' \gets $ \Call{ReplaceNextAngelic}{$P, B$} \Comment{Use $B$ in place of one angelic condition}
                    \State $T' \gets $ \Call{GetPassedTests}{$P',C$}
                    \If{$T' \supseteq T$} \Comment{Successfully resolved that angelic condition}
                        \State $P \gets P'$
                        \State $T \gets T'$
                        \State \emph{success} $\gets \top$ 
                        \State \textbf{break}
                    \EndIf
                \EndRepeat
                \If{$\neg$ \emph{success}}
                    \State \textbf{return} Failure \Comment{Timeout reached}
                \EndIf
            \EndWhile
            \State \textbf{return} $P$ \Comment{All angelic conditions resolved}
        \EndProcedure
    \end{algorithmic}
\end{algorithm}

\subsection{Limitations of Angelic Conditions}
\label{subsec:limitationsAngelic}

Angelic conditions are not very helpful when the desired program's output is a Boolean or small nonnegative integer.
This is because the following angelic programs can pass all test cases for such tasks with an appropriate code path:

\begin{lstlisting}[frame=none, mathescape=true]
boolean unhelpfulBoolean(String str) {
    boolean var = false;
    if ($\angelic$)
        var = true;
    return var;
}

int unhelpfulInteger(String str) {
    int var = 0;
    for (int i = 0; $\angelic$; i++)
        var = i;
    return var;
}
\end{lstlisting}

These angelic programs are almost never correct, leading to wasted effort trying to resolve the conditions. Such tasks
are difficult for \frangel, and for synthesis in general, because the input-output examples provide less information, so
many examples are necessary before the desired behavior becomes apparent and unambiguous.

\section{Experiments}
\label{sec:experiments}

We implemented \frangel\ in about 7500 lines of Java and designed experiments to answer the following questions:
\begin{itemize}[leftmargin=3em]
    \item[\textbf{(Q1)}] How do mining fragments and angelic conditions contribute to \frangel's performance?
    \item[\textbf{(Q2)}] How does \frangel\ compare to previous work?
    \item[\textbf{(Q3)}] How complex are the tasks solved by \frangel?
    \item[\textbf{(Q4)}] How often does \frangel\ produce incorrect programs?
\end{itemize}

We imposed a 30 minute timeout per synthesis task and a 4 GB memory limit for \frangel. Results were averaged over 10
trials, running on Intel Xeon E5-2673 v3 (2.40 GHz) processors.

\subsection{Benchmarks}
\begin{table}
    \caption{Benchmarks used in our experiments. We also list the number of components available to the synthesizer, and
    the number of provided examples, averaged over all tasks contained in each benchmark.}
    \setlength{\tabcolsep}{10pt}
    \begin{tabular} {l c c c}
                                  &       & Avg. \#    & Avg. \#  \\
        Benchmark                 & Tasks & Components & Examples \\
        \toprule
        \emph{Geometry}           & 25    & \phantom{0}400.8      & 4.72     \\
        \emph{ControlStructures}  & 40    & \phantom{0}101.0      & 5.42     \\
        \emph{GitHub}             & 25    & \phantom{0}292.0      & 4.36     \\
        \emph{SyPet}              & 30    & 3639.4     & 1.43     \\
        \bottomrule
    \end{tabular}
    \label{tab:benchmarks}
\end{table}

We evaluated \frangel\ on four benchmarks, comprising 120 tasks in total. Table~\ref{tab:benchmarks} shows the number of
tasks and the average number of components and examples for the benchmarks. 
\begin{itemize}
    \item \emph{Geometry}: We designed 25 tasks of varying difficulty that use a variety of types from the
        \code{java.awt} and \code{java.awt.geom} libraries. These tasks are meant to demonstrate that our approach is
        domain-agnostic, since we use no domain-specific knowledge about these geometry tasks. 12 of these tasks also
        require control structures.
    \item \emph{ControlStructures}: We designed 40 tasks that require using control structures. These tasks were
        designed to cover many common ways of using and combining control structures. Various types are involved,
        including primitives, strings, arrays, and several standard Java collections. Some tasks require combining
        three control structures (sequential and/or nested).
    \item \emph{GitHub}: We selected 25 interesting tasks based on methods from 4 popular Java repositories on
        GitHub. 17 tasks require control structures, and 13 involve custom types defined in the open-source projects.
        Examples were drawn from existing unit tests if available, with some additional examples added to clarify
        functionality and expose special cases where needed.
    \item \emph{SyPet}: We also use the 30 tasks used in the evaluation of \sypet~\cite{sypet}, none of which involve
        control structures. To match \sypet's search space as closely as possible, for these tasks we prevent \frangel\
        from using control structures, primitive operators, literals, and object fields. Although \frangel\ handles
        polymorphism naturally, for this benchmark we also use the same hardcoded polymorphism information used by
        \sypet.
\end{itemize}

We created the \emph{Geometry}, \emph{ControlStructures}, and \emph{GitHub} benchmarks based on the following
general principles:

\begin{itemize}
    \item If \frangel\ is able to infer all necessary classes from the desired method signature, then we did not
        explicitly specify any other classes to use. This occurred for 58 of the 90 new tasks. Otherwise, we specified
        the extra required classes individually if the class containing the desired component(s) is easily identified
        (e.g., \code{java.lang.Math}), or by an entire package (e.g., \code{java.awt.geom.*}) if a specific class would
        not be obvious without knowing the solution.
    \item Examples were generally produced by writing one or two large general-case examples, duplicating them with
        slight modifications, and simplifying them to produce special cases. This frequently involved replacing
        different subsets of the inputs with degenerate values, and/or altering the examples to cover different control
        structure code flows.
    \item Creating examples was a slightly interactive process. By manually observing the fragments mined by \frangel,
        the examples passed, and \frangel's final outputs, we could identify underspecified tasks, incorrect examples,
        and examples that did not cover the special cases as intended. We gradually improved the benchmarks in this
        manner.
    \item We deliberately included tasks that \frangel\ currently fails to solve, so that future work has room to
        demonstrate improvement when using our benchmarks.
\end{itemize}

\subsection{\sypet\ Modifications}

Our experiments compare \frangel\ to \sypet~\cite{sypet}, an existing component-based synthesizer using a datastructure
called a Petri net. We note that \sypet\ does not support control structures, but aside from \sypet, we are not aware of
any other synthesizer that can generate an entire (multi-line) function implementation using arbitrary Java classes.
Hence, we believe \sypet\ to be the existing work most suited to solve our benchmarks.

In order to run \sypet\ on our new benchmarks, we modified the open-source version of \sypet, primarily to provide
support for JAR files compiled with newer versions of Java. As a result, our modified \sypet\ implementation is not
functionally identical to the original open-source version---the modified version does not consider methods and classes
that are deprecated or refer to unavailable classes, effectively pruning the space of programs considered by \sypet,
leading to better efficiency overall. To increase our confidence in our modifications, we ran the modified and original
versions of \sypet\ on \sypet's benchmarks. We observed that the original version solves 25 out of 30 tasks, while our
modified version solves a superset of 26 tasks. Additionally, the modified version is faster for 25 tasks and slower for
1 task, with a median speedup of $1.95\times$.

We also point out that the modified version of \sypet\ is still slower than the times reported in the \sypet\
paper~\cite{sypet} for two reasons. First, the ``synthesis time'' in the \sypet\ paper does not include the time
required to construct an ILP representation of the Petri net and set up a solver for that ILP, a step necessary to find
reachable paths in the Petri net (leading to solution sketches). Our measurements of \sypet's runtime do include this
step.\footnote{As in the \sypet\ paper, we exclude the time used by \sypet\ to compile candidate programs. But to be
conservative in our claims, \frangel's equivalent (interpreting candidates via reflection without compilation) is
\emph{included} in our reported times.} Second, the \sypet\ paper reports times for a closed-source version of \sypet\
using some optimizations not included in the open-source version. Because we had to modify \sypet\ to address the above
issues, we extended the open-source version without those optimizations.

For the rest of the discussion, we use ``\sypet'' to refer to our modified version of \sypet. When running \sypet\ on
our new benchmarks, we provided \sypet\ with all of the constants and operators that are used by \frangel\ (by wrapping
them inside functions that \sypet\ can call). \sypet\ and \frangel\ were also given the same libraries (i.e., Java
classes). Because \sypet\ is deterministic and memory-intensive, we ran it once on each task with 12 GB of
memory.\footnote{Our experiments used machines with only 14 GB of RAM total. Experiments in the \sypet\ paper used 32 GB
of memory.} Even so, \sypet\ still crashed from insufficient memory on some tasks.

\subsection{Results and Discussion}
\label{subec:results}

\begin{figure}
    \centering
    \includegraphics[width=\linewidth]{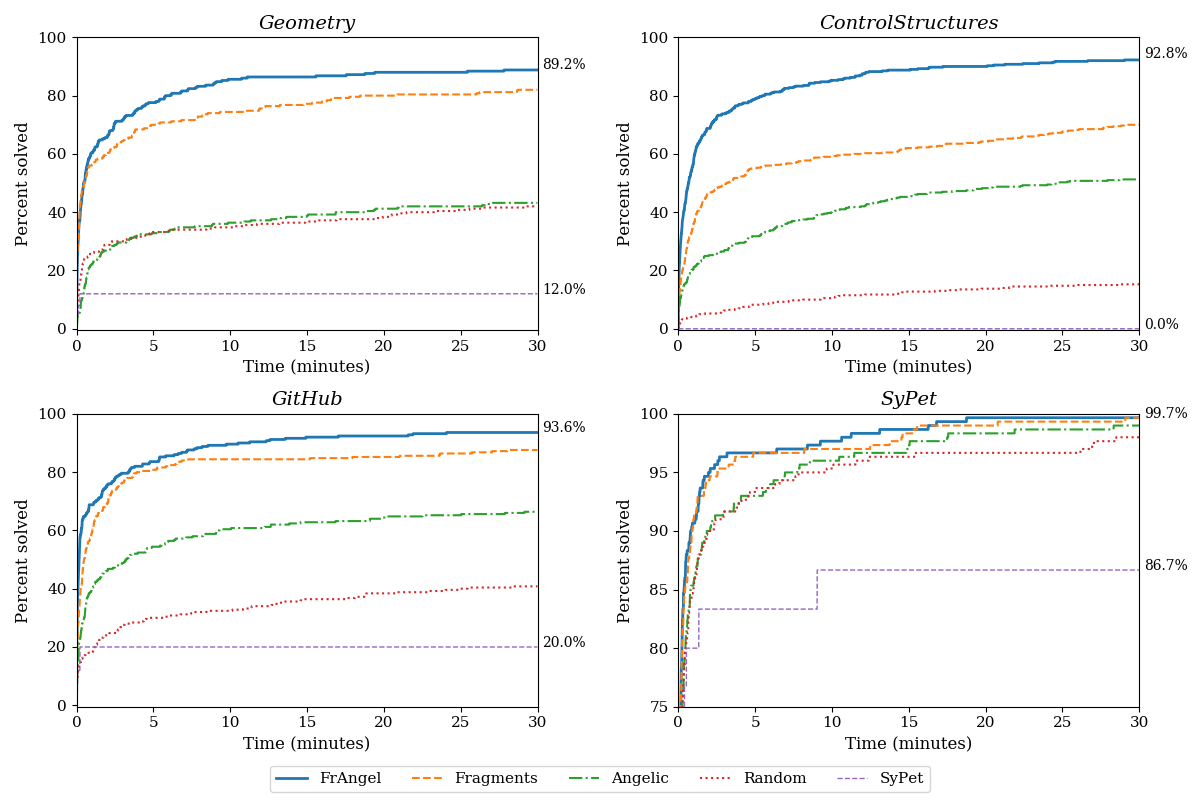}
    \caption{Results for the four benchmarks. The success rates at 30 minutes for \frangel\ and \sypet\ are listed on
    the right of each plot. (The plot for the \emph{SyPet} benchmark uses a different \emph{y}-axis scale.)}
    \label{fig:results}
\end{figure}

Using our benchmarks, we evaluated \frangel\ and \sypet, together with versions of \frangel\ that that only mine
fragments (``Fragments'') or only use angelic conditions (``Angelic''), or neither (``Random,'' a na\"ive random
search). The results are shown in Figure~\ref{fig:results}, which plots the time required (per task) to solve a fraction
of the benchmarks, e.g., 92.8\% of tasks in \emph{ControlStructures} were solved by \frangel\ before the 30 minute
per-task timeout. The plots show that Random outperforms \sypet, Fragments and Angelic are generally better than
Random, and \frangel\ performs the best.

To illustrate \frangel's abilities, in Figure~\ref{fig:interestingResults} we list some of the most interesting programs
generated (we removed unnecessary braces for space). In Figure~\ref{subfig:eccentricity}, \frangel\ calculates ellipse
eccentricity in a way simpler than the authors' best handwritten solution, implicitly taking advantage of the identity
$\cos(\sin^{-1}(x)) = \sqrt{1 - x^2}$. In Figure~\ref{subfig:sort}, \frangel\ implements a sorting algorithm in only 30
seconds when given a \code{swap} function.

Additionally, Figure~\ref{fig:fastResults} shows results for several tasks that can be solved in under 10 seconds,
showing that \frangel\ is fast enough to be used interactively for the easier but still interesting tasks in our
benchmarks. Note that some of the implementations are not the most natural. For instance, in
Figure~\ref{fig:fastResults}'s \code{longestString} implementation, \code{elem1.startsWith("", var1.length())} is more
naturally written as \code{var1.length() <= elem1.length()}, although both have the same AST size. This is a fundamental
limitation of the synthesis setting, where there might be many correct ways of obtaining the desired functionality, and
a random search-based synthesizer is not at all guaranteed to find the ``best'' implementation.

\begin{figure}
    \centering
    \begin{subfigure}[b]{\linewidth}
        \begin{lstlisting}
// Time: 280.0 sec (10/10), Size: 15, Components: 211, Examples: 8, Benchmark: Geometry
// Fragments: 285.6 sec (10/10), Angelic: not solved (0/10), Random: not solved (0/10)
static double ellipseEccentricity(Ellipse2D ellipse) {
    double var1 = 0.0;
    var1 = ellipse.getHeight() / ellipse.getWidth();
    return Math.cos(Math.asin(Math.min(var1, 1.0 / var1)));
}       \end{lstlisting}
        \vskip -8pt
        \caption{\frangel's approach to computing ellipse eccentricity is actually simpler than our own handwritten
        solution, which used the formula $e = \sqrt{1 - b^2/a^2}$, where $a$ and $b$ are the lengths of the major and
        minor axes respectively. Mining fragments from special-case programs is a successful strategy for this task.}
        \label{subfig:eccentricity}
    \end{subfigure}
    \vskip 8pt
    \begin{subfigure}[b]{\linewidth}
        \begin{lstlisting}
// Time: 30.1 sec (10/10), Size: 21, Comp.: 37, Examples: 5, Bench.: ControlStructures
// Fragments: not solved (0/10), Angelic: not solved (0/10), Random: not solved (0/10)
static void sortArrayGivenSwap(double[] arr) {
    for (int i1 = 0; i1 < arr.length; i1++)
        for (int i2 = 0; i2 < i1; i2++)
            if (arr[i1] < arr[i2])
                Swap.swap(arr, i2, i1);
} 	\end{lstlisting}
        \vskip -8pt
        \caption{If given a \code{swap} function, \frangel\ can use nested control structures to implement a variation
        of Insertion Sort in 30 seconds. In most other runs, \frangel\ produces a Bubble Sort implementation instead.}
        \label{subfig:sort}
    \end{subfigure}
    \vskip 6pt
    \begin{subfigure}[b]{\linewidth}
        \begin{lstlisting}
// Time: 980.5 sec (7/10), Size: 17, Comp.: 36, Examples: 5, Bench.: ControlStructures
// Fragments: not solved (0/10), Angelic: 655.6 sec (10/10), Random: not solved (0/10)
static LinkedListNode reverseLinkedList(LinkedListNode node) {
    LinkedListNode var1 = null;
    while (!(null == node)) {
        var1 = new LinkedListNode(node.getValue(), var1);
        node = node.getNext();
    }
    return var1;
} 	\end{lstlisting}
        \vskip -8pt
        \caption{\frangel\ manipulates multiple variables to reverse a linked list. Angelic conditions are crucial for
        this task.}
    \end{subfigure}
    \vskip 6pt
    \begin{subfigure}[b]{\linewidth}
        \begin{lstlisting}
// Time: 432.1 sec (10/10), Size: 15, Components: 165, Examples: 2, Benchmark: GitHub
// Fragments: 196.8 sec (10/10), Angelic: not solved (0/10), Random: not solved (3/10)
static IndexResponse elasticsearch_fromXContent(XContentParser parser) {
    IndexResponse.Builder var1 = new IndexResponse.Builder();
    parser.nextToken();
    while (XContentParser.Token.FIELD_NAME.equals(parser.currentToken())) {
        IndexResponse.parseXContentFields(parser, var1);
        parser.nextToken();
    }
    return var1.build();
}       \end{lstlisting}
        \vskip -8pt
        \caption{\frangel\ handles complex custom types in a domain-agnostic way. In this task, \frangel\ automatically
        identifies all necessary methods and classes from the function signature without user assistance.}
    \end{subfigure}
    \vskip 8pt
    \caption{Some of the most interesting programs synthesized by \frangel. We display the solution produced by the
    6th-fastest solve out of 10 runs. We also list the size of the displayed program, the number of components and
    examples for the task, and the 6th-fastest solve time and success rate for all \frangel\ variants.}
    \label{fig:interestingResults}
\end{figure}

\begin{figure}
    \centering
    \begin{lstlisting}
// Time: 3.7 sec, Size: 12, Components: 191, Examples: 4, Benchmark: Geometry
static Rectangle2D.Double rectangleUnion(Rectangle2D[] rects) {
    Rectangle2D.Double var1 = new Rectangle2D.Double();
    var1.setRect(rects[0]);
    for (Rectangle2D elem1 : rects)
        var1.add(elem1);
    return var1;
}

// Time: 9.1 sec, Size: 6, Components: 1421, Examples: 2, Benchmark: Geometry
static void rotatePointDegrees(Point2D point, double degrees) {
    AffineTransform.getRotateInstance(Math.toRadians(degrees)).transform(point, point);
}

// Time: 5.2 sec, Size: 13, Components: 137, Examples: 7, Benchmark: ControlStructures
static String longestString(List<String> list) {
    String var1 = "";
    for (String elem1 : list)
        if (elem1.startsWith("", var1.length()))
            var1 = elem1;
    return var1;
}

// Time: 9.2 sec, Size: 15, Components: 48, Examples: 6, Benchmark: ControlStructures
static void rotateQueue(Queue<Object> queue, int amount) {
    if (!queue.isEmpty())
        for (int i1 = 0; i1 < amount % queue.size(); i1++)
            queue.add(queue.poll());
}

// Time: 5.9 sec, Size: 14, Components: 119, Examples: 4, Benchmark: GitHub
static String guava_getPackageName(String classFullName) {
    if (!classFullName.contains("."))
        classFullName = ".";
    return classFullName.substring(0, classFullName.lastIndexOf("."));
}

// Time: 0.8 sec, Size: 11, Components: 213, Examples: 4, Benchmark: GitHub
static void zxing_maybeAppend(String[] values, StringBuilder result) {
    if (false == (values == null))
        for (String elem1 : values)
            ParsedResult.maybeAppend(elem1, result);
}

// Time: 8.0 sec, Size: 5, Components: 1093, Examples: 2, Benchmark: SyPet
static double[] sypet_06_solveLinear(double[][] mat, double[] vec) {
    return MatrixUtils.inverse(new BlockRealMatrix(mat)).operate(vec);
}

// Time: 0.3 sec, Size: 7, Components: 6072, Examples: 1, Benchmark: SyPet
static DocumentType sypet_26_getDoctypeByString(String xmlStr) {
    return DocumentBuilderFactory.newInstance().newDocumentBuilder().parse(
            new InputSource(new StringReader(xmlStr))).getDoctype();
}   \end{lstlisting}
    \vskip -3pt
    \caption{
        \frangel\ solves a variety of interesting tasks in under 10 seconds, making the tool suitable for interactive
        use. We display the results and times for the 6th-fastest solve out of 10 runs.
    }
    \label{fig:fastResults}
\end{figure}

\paragraph{Mining fragments, angelic conditions.} To answer \textbf{(Q1)}, we compare \frangel\ to Fragments, Angelic,
and Random. Based on the relative performance of these variants, we conclude that mining fragments is an important
strategy in all of our benchmarks. However, mining fragments is not as useful for the \emph{SyPet} benchmark, because
most of those tasks are not special-case similar or do not expose any special cases in the provided examples. Of the 30
tasks in the \emph{SyPet} benchmark, 19 only provide a single example (test case). In this situation, mining fragments
has no effect at all, since we only mine fragments from programs that pass a nonempty subset of test cases, and the
search would terminate successfully as soon as the single test case is passed.

\begin{table}
    \caption{Various statistics for \frangel\ related to mining fragments (over all successful runs, excluding the
    \emph{SyPet} benchmark). Roughly speaking, the last column measures the average amount of overlap between mined
    fragments and the final program.}
    \setlength{\tabcolsep}{10pt}
	\begin{tabular} {l c c c}
	           & \# Programs              & \# Fragments              & Avg. Fragment \\
	           & Remembered               & Mined                     & Usefulness \\
	    \toprule
        Mean   & \phantom{0}6.4           & \phantom{0}19.4           & \phantom{0}62.2\% \\
        Median & \phantom{0}4\phantom{.0} & \phantom{0}16\phantom{.0} & \phantom{0}63.9\% \\
        Max    & 62\phantom{.0}           & 148\phantom{.0}           & 100.0\% \\
	    \bottomrule
	\end{tabular}
    \label{tab:fragmentStats}
\end{table}

Table~\ref{tab:fragmentStats} lists various statistics about \frangel\ related to mining fragments, considering all
tasks except for the \emph{SyPet} benchmark. For most tasks, \frangel\ only needs to mine fragments from a handful of
programs. Note that the number of remembered programs is much smaller than the theoretical maximum, i.e., one remembered
program for each subset of test cases. Our tasks have up to 12 examples, yet the number of programs remembered is always
relatively small.

For a given mined fragment, its ``usefulness'' measures the maximal amount of overlap between that fragment and some
fragment in the final program. For example, if the final program is \mbox{\code{a + b.foo(c)}}, then the fragment
\code{b.foo(1)} has a usefulness of 67\%, since 2 AST nodes (i.e., \code{b} and \code{foo}), out of 3 total, match the
fragment \code{b.foo(c)} from the final program.

For each successful run of \frangel, we average the usefulness of all mined fragments to obtain the data points
summarized in the table. Hence, in the average run of \frangel, the average mined fragment has at most a 62.2\% overlap
with some fragment in the final program. This provides empirical evidence that mined fragments are useful for synthesis.

\begin{figure}
    \centering
    \includegraphics[width=\linewidth]{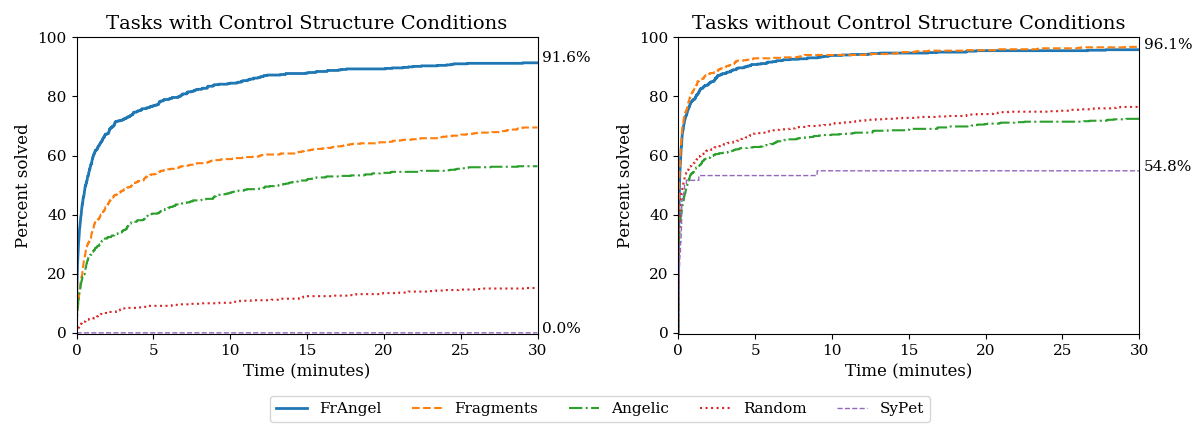}
    \caption{Results separated by whether the task involves control structures with conditions (i.e., conditionals and
    loops excluding for-each loops). The success rate at 30 minutes for \frangel\ and \sypet\ are listed on the right of
    each plot.}
    \label{fig:conditions}
\end{figure}

Angelic conditions are also generally helpful, but their effect is most prominent in the \emph{ControlStructures}
benchmark, where all tasks require control structures. This is expected, since angelic conditions are only relevant
when control structures with conditions are involved (i.e., not for-each loops). This distinction is more apparent in
Figure~\ref{fig:conditions}, which shows that angelic conditions are very helpful when control structures with
conditions are involved, while the strategy causes a slight slowdown for tasks without control structure conditions.

\paragraph{\sypet\ comparison.} We address \textbf{(Q2)} by comparing \frangel\ to the previous work \sypet\ in
Figure~\ref{fig:results}. It might seem surprising that a na\"ive random search (i.e., Random) outperforms \sypet. One
contributing factor is that our grammar is more expressive than \sypet's grammar. For instance, Random can use control
structures but \sypet\ cannot, which explains why \sypet\ is not able to solve any tasks in our \emph{ControlStructures}
benchmark.

We also note that \sypet\ is designed for tasks that involve many datatype conversions. However, most of our new
benchmarks perform many operations on the same types, e.g., \code{String} operations that return another \code{String}.
We observed that \sypet\ struggles with such tasks. In particular, suppose \code{a}, \code{b}, \code{c}, and \code{d}
are \code{String}s. \sypet\ takes 0.29 seconds to synthesize \code{a + b}, 28.2 seconds to synthesize \code{a + b + c},
and over 30 minutes (i.e., timeout) to synthesize ${\code{a + b + c + d}}$. In contrast, Random takes 0.003, 0.107, and
11.4 seconds respectively to synthesize those expressions.

To explain why Random outperforms \sypet\ even on the \emph{SyPet} benchmark, we point out that our implementation
achieves a very high throughput, defined as the number of programs tried per unit time. On the \emph{SyPet} benchmark,
\sypet\ executes on average 37 programs per second (PPS) on successful runs, not counting the time used to compile
programs. In contrast, Random's average throughput exceeds 14,000 PPS. This difference is partially because \sypet\
spends a large portion of time constructing Petri nets and solving ILP and SAT problems, while Random does nothing
beyond generating and executing candidate programs. On the \emph{SyPet} benchmark, the other \frangel\ variations have a
similar throughput. Note that angelic conditions have no effect because, for consistency in our comparison to \sypet, we
do not use control structures for the \emph{SyPet} benchmark.

On our new benchmarks (i.e., excluding the \emph{SyPet} benchmark), Random achieves a higher average throughput of
31,400 PPS because many tasks involve simpler types, so candidate programs can be evaluated much more quickly. \frangel\
spends more time with each program---angelic programs involve executing many code paths, resolving conditions takes
time, and mining fragments requires checking all test cases (whereas Random can break early as soon as one test case
fails). Thus, \frangel\ only tries on average 5,300 PPS. Fragments and Angelic have throughputs of 17,700 and 5,600 PPS
respectively.

In our opinion, one takeaway from \frangel's comparison to \sypet\ is that lightweight heuristic approaches (e.g.,
\frangel\ or even Random) can be surprisingly effective in domains where execution of candidate programs is cheap,
compared to other heavyweight approaches (e.g., using symbolic reasoning or deep learning) that can be fundamentally
limited in throughput.

\begin{wrapfigure}[15]{R}{0.5\textwidth}
    \centering
    \vspace{-12pt}
    \includegraphics[width=0.5\textwidth]{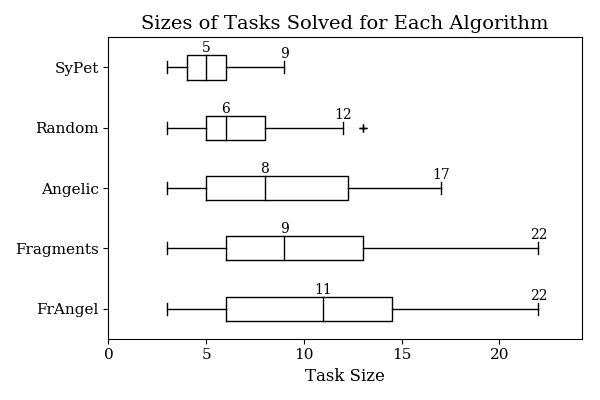}
    \caption{Sizes of the tasks solved for each algorithm. A \frangel\ variant ``solves'' a task if it succeeds at least
    6 times out of the 10 runs.}
    \label{fig:sizes}
\end{wrapfigure}

\paragraph{Task complexity.} To address \textbf{(Q3)}, we use program size (number of AST nodes) as a measure of the
underlying task's difficulty. More precisely, for each task, we consider all programs solving the task (produced by any
\frangel\ variant among all 10 runs), and we define that task's ``size'' to be the minimum size of such a program. It
is possible that a task could be solved with a program of even smaller size; our notion of task size is only an
approximate measure of the task's difficulty.

Figure~\ref{fig:sizes} shows the sizes of the tasks solved by each algorithm, where we consider a \frangel\ variant to
``solve'' a task if it succeeds at least 6 times out of the 10 runs. Compared to the other methods, \frangel\ solves the
most difficult tasks, with a median task size of 11 and maximum of 22. \sypet\ is only able to solve tasks with size
less than 10, most of which are from the \emph{SyPet} benchmark. This suggests that the tasks in our new benchmarks are
much more difficult than the tasks in \emph{SyPet}.

Note that Figure~\ref{fig:sizes} does not reflect programs that are larger than necessary. On average, a program
produced by \frangel\ has a size $1.16\times$ the task's size, and 64\% of the time, the two sizes are equal. We thus
conclude that \frangel\ is reasonably consistent in the sizes of its solutions.

\begin{figure}
    \centering
    \begin{lstlisting}
static int sypet_23_getOffsetForLine(Document doc, int line) {
    int var1 = 0;
    line = doc.getDefaultRootElement().getElement(var1).getEndOffset();
    return line;
}
    
static double distanceInCircle(Point2D point, Ellipse2D circle) {
    if (!circle.contains(point))
        point = new Point2D.Double(Double.MAX_VALUE, 0.0);
    return point.distance(circle.getCenterX(), circle.getCenterY());
}   \end{lstlisting}
    \caption{
        Two representative false positives produced by \frangel. (1) The first program returns the end offset of the element
        at index \code{0}, but it should return the start offset of the element at index \code{line}. \sypet's original
        benchmark only provides one example for this task, where \code{line} is 1. Further examples with different
        values of \code{line} would rule out this false positive. (2) \code{distanceInCircle} is supposed to return the
        distance between \code{point} and the center of \code{circle} if \code{point} is contained inside \code{circle},
        or \code{Double.POSITIVE\_INFINITY} otherwise; this implementation is incorrect in the latter case when
        \code{circle.getCenterX()} is \code{Double.MAX\_VALUE}.
    }
    \label{fig:falsePositive}
\end{figure}

\paragraph{False positives.} To answer \textbf{(Q4)}, we manually verified (by inspection and further testing) a random
selection of 100 programs out of 1127 produced by \frangel.\footnote{The programs and our verdicts can be found at
\url{https://www.github.com/kensens/FrAngel}~\cite{frangel-url}.} We found that 7 out of the 100 programs were
\emph{false positives}, i.e., they pass all given test cases but are not completely correct. Two representative false
positives with explanations are shown in Figure~\ref{fig:falsePositive}.

The first false positive is for a \emph{SyPet} task that only has one example. Hence, the task is somewhat ambiguously
specified, so it is expected that random search approaches like \frangel\ may produce false positives.\footnote{\sypet\
is guaranteed to solve the task correctly because it is deterministic and because the \sypet\ authors added examples
until \sypet\ produced the expected result.} Further examples, if chosen reasonably, would eliminate that false
positive. Of the 7 false positives, 3 were for \emph{SyPet} tasks with only one example.

The second false positive is more difficult to guard against, since it is correct except for a narrow class of inputs.
Furthermore, there are multiple variations of the false positive, and ruling them out via examples would require a
different specially-crafted input for each variation.

Due to the possibility of false positives, we stress that the human users should independently verify the correctness of
programs produced by \frangel. Note that the possibility of false positives is common to most example-based synthesis
approaches.

\section{Related Work}

Past efforts in program synthesis have been quite diverse. Some approaches search through a space of programs using
brute force~\cite{ssle, mutual}, MCMC search~\cite{stoke, mimic}, genetic programming~\cite{genetic}, or neural
networks~\cite{deepcoder, robustfill}. Others use constraint-solving~\cite{oracle, slfp, sypet, verification, morpheus}
or version space algebras~\cite{vsa, flashfill}. \frangel's random search is most similar to the MCMC and genetic
programming approaches, and \frangel\ enjoys the simplicity and generality offered by random search. However, unlike
many solver-based approaches, \frangel\ cannot guarantee program correctness.

\frangel's focus on finding distinct program behaviors is one quality that sets it apart from approaches using other
similar synthesis paradigms. For instance, standard enumerative approaches and MCMC search do not attempt to combine
functionality drawn from previous partially-successful programs, which \frangel\ does naturally by using mined code
fragments. Furthermore, the candidate programs produced by enumeration or MCMC are often highly correlated, which can
lead to a loss in diversity. On the other hand, \frangel's use of fragments helps to balance diverse samples in the
program space with learning from experience.

Genetic programming is based on the idea of combining previous successful programs to create better ones. However, the
approach typically optimizes for a metric such as the number of test cases passed. This objective can lead to
remembering many equally-high-scoring programs with nearly the same behavior, while programs containing desirable
edge-case functionality might be discarded if they pass fewer tests. In contrast, \frangel\ explicitly remembers the
simplest program for every behavior distinguishable using the test cases, leading to a greater diversity in remembered
program behaviors. \frangel's strategy cannot be emulated within the standard genetic programming framework, as an
equivalent ``fitness function'' must depend on the whole population: whether a program remains in the population depends
on the presence of a simpler program passing the same tests or a superset of them. Furthermore, mining fragments can be
viewed as a generalization of crossover and mutation from genetic programming, since \frangel\ can combine and modify
fragments from several programs simultaneously.

\subsection{Synthesis Setting}

\paragraph{Component-based synthesis} Component-based synthesis approaches are interesting because they generalize to
different user-provided components. Such approaches have been applied to many areas, including string
manipulation~\cite{mlpbe}, bitvector algorithms~\cite{slfp}, deobfuscation~\cite{oracle}, geometric
constructions~\cite{geometry}, and transformations of data~\cite{hades, lambda2, morpheus, progfromex}. Some works focus
on code completion, finding compositions of components that typecheck given the surrounding code context~\cite{jungloid,
slang, insynth, insynth2, codehint}.

From a user-interface perspective, our work is most similar to \sypet~\cite{sypet}. Like \frangel, \sypet\ is a
component-based synthesizer that synthesizes Java programs from examples using arbitrary libraries. \sypet\ uses a Petri
net datastructure to find sequences of function calls that produce the desired output type. However, \sypet\ cannot
synthesize control structures and struggles with many operations on the same types, while \frangel\ excels in these
areas. Our experimental results show that \frangel\ outperforms \sypet\ even on \sypet's benchmarks.

CodeHint~\cite{codehint} is another component-based synthesis tool that produces a ranking of Java expressions given the
surrounding code context. It performs synthesis at run-time, i.e., inside a debugging session paused at a breakpoint,
using the actual execution context. CodeHint only produces one-line expressions, while most of our tasks require the use
of multiple statements.

\paragraph{Programming by example (PBE)} Many synthesizers including \frangel\ use input-output examples to specify the
desired behavior~\cite{flashfill, sypet, hades, lambda2, morpheus, progfromex, ted, tds, mlpbe}. Most of these focus on
a particular domain and require domain-specific knowledge in the form of a DSL or carefully-chosen components. In
contrast, \sypet\ and \frangel\ do not require domain-specific knowledge, only a list of relevant libraries.

A PBE approach by \citet{mlpbe} uses examples to extract ``features'' for string-processing tasks. For
instance, if the input is a list with duplicate strings, and the output is a list without duplicates, this feature
suggests using a \code{dedup} operator. This is similar to \frangel's technique of mining fragments, but \frangel\
obtains fragments from previous programs, not by extracting features from examples.

\subsection{Synthesis Techniques}

\paragraph{Learning from previous programs.} LaSy~\cite{tds} is a PBE approach inspired by test-driven development. From
a sequence of increasingly-complex test cases, LaSy synthesizes a sequence of programs, each a modification of the
previous program, where program $i$ passes the first $i$ test cases. Our idea of mining fragments (i.e., by remembering
the simplest program that passes each subset of tests) could be seen as a generalization of the LaSy approach. In
particular, we do not assume that test cases are ordered by complexity; instead, we allow remembered programs and
fragments to build upon each other in a nonlinear fashion---new programs can use fragments extracted from any
combination of previously-remembered programs.

\textsc{Strata}~\cite{strata} synthesizes formal semantics of x86-64 instructions and uses previously-learned semantics
to bootstrap the learning process of unknown semantics. Note that the problem addressed by \textsc{Strata} is very
different from ours; \frangel\ synthesizes the source code of Java functions instead of assembly instruction semantics.

\paragraph{Program sketching.} A common synthesis technique, popularized by \textsc{Sketch}~\cite{sketch}, is to provide
the overall code context (a program \emph{sketch}) while leaving ``holes'' to be filled by the
synthesizer~\cite{angelic, sypet, lambda2, morpheus, slang, codehint}. This technique is frequently used by synthesizers
including \frangel\ to perform synthesis in two steps: first identify promising program sketches and then complete the
sketches. In \frangel, angelic conditions are holes in place of control structure conditions, so angelic programs are a
kind of program sketch.

\frangel's angelic conditions are related to a programming model called \emph{angelic nondeterminism}~\cite{angelic}. In
this paradigm, the programmer first writes a program sketch using a \code{choose} operator, which nondeterministically
evaluates to a constant such that execution passes all \code{assert} statements (if possible). Thus, programs with
\code{choose} operators are similar to our angelic programs. The programmer can then iteratively refine the program,
slowly removing \code{choose} operators while maintaining the program's validity. Eventually, when all \code{choose}
operators are removed, the programmer arrives at a deterministic program. This is similar to our approach of resolving
angelic conditions one at a time. While the ideas in \frangel\ overlap with those in angelic nondeterminism, they have
different purposes: \frangel\ uses angelic conditions to decompose a random search by evaluating the quality of partial
programs, while angelic nondeterminism uses the \code{choose} operator to represent a class of programs that is
gradually narrowed down by the user.

\paragraph{Synthesizing control structures.} As noted in Section~\ref{sec:intro}, many component-based synthesizers
cannot handle control structures, especially loops~\cite{sypet, slfp, oracle, mlpbe}. Beyond component-based synthesis,
some approaches in the wider program synthesis literature do handle loops~\cite{mimic, msee, simd, verification,
mutual}, but many do not~\cite{stoke, deepcoder, robustfill, eusolver, vsa}. Hence, handling combinations of control
structures is a distinguishing aspect of \frangel.

\frangel\ uses the idea of generating and evaluating the body of a control structure before learning the condition. This
provides a decomposition of the synthesis task, ultimately allowing \frangel\ to synthesize programs using combinations
of loops and conditionals. The concept of leaving a control structure condition unspecified has been applied to
\code{if} and \code{switch} statements by several other works~\cite{mimic, flashfill, eusolver, hades}. This is often
presented as learning several programs and then learning the classifier that chooses which program to use for a given
input. Applying this to loops, as done by \frangel, requires nontrivial extensions such as the partial enumeration
procedure in Section~\ref{subsec:angelicExecution}.

The \textsc{Mimic} algorithm~\cite{mimic} infers a loop structure by analyzing execution traces (i.e., sequences of
memory reads and writes). \textsc{Mimic} uses MCMC search and requires an executable that can produce traces on new
inputs. While \textsc{Mimic} and \frangel\ both try to learn control structures, their settings differ since \frangel\
does not require execution traces.

Some synthesizers handle loops with techniques from program verification~\cite{verification, simd, mutual}, but such
approaches have only been demonstrated in low-level code (i.e., only using primitive operations and not library
functions). Other approaches call components that loop internally~\cite{msee, progfromex, lambda2, hades} or use a DSL
for specialized loops~\cite{flashfill, tds}, but these are only applicable in specific domains where the loops can be
categorized into a few common types. \frangel\ is the first approach to our knowledge that combines synthesis of control
structures and domain-agnostic library function calls.

\section{Conclusion}

\frangel\ is a domain-agnostic approach to component-based program synthesis from examples that mines code fragments
from partial successes and uses angelic conditions to evaluate control structure sketches. These techniques help
\frangel\ discover programs with new and complex behaviors related to the synthesis task. Our experiments show that
\frangel\ can generate interesting programs using various libraries and combinations of control structures within
several seconds.

For future work, we note that most of \frangel's randomly-generated candidate programs would seem unnatural to human
programmers, with symptoms including code with no effect (e.g., \code{str.length();}), excessive complexity (e.g.,
\code{rect.getBounds2D().getBounds2D()} when \code{rect} suffices), and improper use of variables (e.g., assigning to a
variable that is never used again). While \frangel\ can afford to process unnatural candidate programs due to its high
throughput, the quality of the candidate programs could be greatly improved. This could allow \frangel\ to use local
variables in more complex ways, which is an aspect of programming that \frangel\ currently struggles with. Furthermore,
when generating code with fragments, we uniformly sample from all mined fragments. One could imagine a more intelligent
scheme that takes into account information such as the fragment's size and the test cases passed by remembered programs
containing it.

As another perspective on our work, \frangel\ demonstrates that there is rich information in well-chosen test cases that
can substantially aid program synthesis. In particular, \frangel\ uses simple and varied test cases to give ``partial
credit'' to simplifications of the desired task and to efficiently evaluate the feasibility of angelic programs. More
generally, one could imagine other partial credit schemes, possibly with higher granularity, that convey even richer
information. This could be a key component in scaling synthesis to more complex programs.

\begin{acks}
    The authors would like to thank Rishabh Singh and Osbert Bastani for their insightful discussions and assistance;
    Mina Lee, Cynthia Yin, and the anonymous reviewers for their valuable feedback on the paper; and Cynthia, Evan Liu,
    and Eva Zhang for their help in testing \frangel's user interface.
\end{acks}

\bibliography{./paper}

\clearpage
\appendix
\section*{Appendix}

Some minor details of our \frangel\ implementation are described below.

\subsection*{Program Interpreter}

Compiling Java code is quite expensive. To evaluate candidate programs as quickly as possible, \frangel\ instead
interprets programs using Java reflection. The interpreter accepts a Java program (as an AST, restricted to the grammar
in Figure~\ref{fig:grammar}), inputs, and expected outputs, and returns whether or not the program behaves as expected.
If the program uses angelic conditions, the interpreter also accepts a code path and returns the actual code path (as
described in Section~\ref{sec:angelic}). The interpreter terminates execution early if the run consumes too much time or
memory, implemented using limits on the number of loop iterations and sizes of objects.

In rare scenarios, \frangel's interpreter will invoke a method with inputs that cause it to loop indefinitely. The
interpreter is currently not equipped to deal with such cases (requiring the \frangel\ process to be restarted),
although further engineering effort using multithreading or multiprocessing should allow the interpreter to forcibly
terminate execution of methods based on elapsed time.

\subsection*{Skipping and Pruning Programs}

If \frangel\ generates a non-angelic candidate program that it has already seen, evaluation of the program is skipped
since it cannot provide new information. If \frangel\ generates a duplicate angelic program, its evaluation is skipped
with $\frac{3}{4}$ probability (resolving conditions is a random process that might or might not succeed, so \frangel\
re-evaluates with probability $\frac{1}{4}$ for robustness). These checks are performed efficiently with low memory
overhead by compressing every program into a compact and unique String encoding and storing the encodings in hashtables.
One million encodings can be stored in about 200 MB of memory. We bound the overall memory usage of \frangel\ by
limiting the number of stored encodings. In our experiments, we store at most 5 million encodings each for angelic and
non-angelic programs.

In addition, \frangel\ prunes away control structures with non-angelic conditions that do not involve any variables
(i.e., the condition is effectively constant). Furthermore, if any line in the program is too large, \frangel\ skips the
program entirely. This prevents \frangel\ from synthesizing code that would be too convoluted for a human to easily
parse.

\end{document}